\documentclass[twocolumn,a4paper]{autart}    % Enable this line and disable the 
\usepackage[bookmarks=false]{hyperref}
\usepackage{graphicx}
\usepackage{amsmath,amssymb,amscd,amsfonts}
\usepackage{bm}
\usepackage{float,mathrsfs}
\usepackage{threeparttable}
\usepackage{color}
\usepackage{stfloats}
\usepackage{subfigure}

\newtheorem{theorem}{Theorem}
\newtheorem{proposition}{Proposition}
\newtheorem{lemma}{Lemma}
\newtheorem{definition}{Definition}

\newtheorem{assumption}{Assumption}

\begin{document}

\begin{frontmatter}

\title{A two-disk approach to the synthesis of coherent passive equalizers for linear quantum systems} % Title, preferably not more 

\thanks[footnoteinfo]{This work was supported by the Australian Research
Council under the Discovery Projects funding scheme (project
DP200102945).}

\author[1,a]{Valery Ugrinovskii},
\author[1,2,b]{Shuixin Xiao}\ead{shuixin.xiao@anu.edu.au} 
\thanks[a]{Valery Ugrinovskii sadly passed away suddenly. This manuscript is a result of his work and dedication.}
\thanks[b]{Corresponding author.}

\address[1]{School of Engineering and Technology, University of New South Wales, Canberra ACT 2600, Australia}                                           
\address[2]{School of Engineering, Australian National University, ACT 2601, Australia}

\begin{keyword}                            
Coherent equalization, robust control, quantum control, linear quantum systems       
\end{keyword}

\begin{abstract}                          % Abstract of not more than 200 words.
The coherent equalization problem consists in designing a quantum system acting as a mean-square near-optimal filter for a given quantum communication channel. The paper develops an improved method for the synthesis of transfer functions for such equalizing filters, based on a linear quantum system model of the channel. The method draws on a connection with the two-disk problem of ${H}_{\infty}$ control for classical (i.e., non-quantum) linear uncertain systems. Compared with the previous methods, the proposed method applies to a broader class of linear quantum communication channels.
\end{abstract}

\end{frontmatter}
\endNoHyper
\section{Introduction}
\begin{figure}
	\centering
	\includegraphics[width=1\columnwidth]{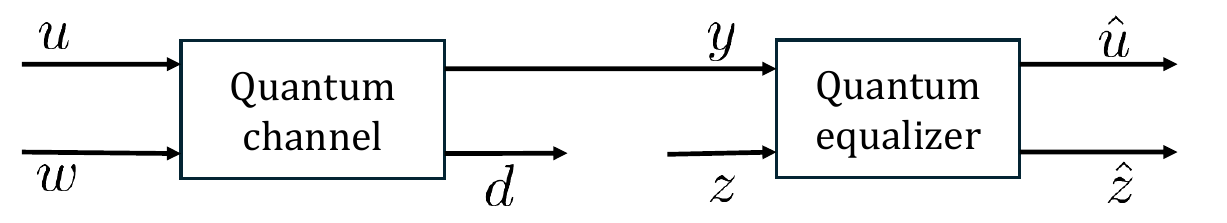}
	{\caption{A general quantum equalization system; e.g., see \cite{Ugrinovskii2024}.}\label{fig1}}
\end{figure}

The coherent equalization problem consists in designing a quantum system acting as a filter for a given quantum communication channel. The concept of equalization is illustrated in Fig.~\ref{fig1}. Both the channel and the equalizer are linear open quantum systems. The channel system interacts with an input Gaussian field. The first $n$ modes of this field (symbolized as $u$ in the figure) are engineered to carry a message transmitted over the channel. The remaining modes (symbolized as $w$) describe the system environment. Interactions with the channel system distort the transmitted signal. To mitigate these distortions, another quantum system is introduced, called the quantum equalizer. This system must be designed so that when it interacts with the channel output field and its own environment, the field resulting from these interactions matches closely (in the mean-square sense) the driving field of the channel.

A similar equalization problem is well known in the classical (i.e., non-quantum) communications. It is concerned with compensating signal degradations caused by the channel due to its limited bandwidth and noise. The initial solution to this problem is due to N. Wiener who showed that channel equalization can be cast as minimization of the error covariance between the channel stationary input signal and the equalizer output signal. Recent references \cite{valcdc,valcdc2,Ugrinovskii2024,Ugrinovskii2024a} extended this approach to quantum systems, with a notable
additional requirement that the optimal (near-optimal) equalizer must be realizable as a physical quantum system; hence the name \emph{coherent equalization}. 
Quantum communication over Gaussian and more general bosonic channels has also been extensively studied in the literature~\cite{g11,g1,g2,g3,g5}. Several works have also investigated the communication capacity of such channels using an amplifier~\cite{g2,g5}. For instance, Ref.~\cite{g2} employs an amplifier to compute the ultimate capacity of bosonic channels, where the amplifier plays a role similar to our quantum equalizer in Fig.~\ref{fig1}, though with a different objective of maximizing the Holevo information under an average photon-flux constraint. Additionally, related studies on quantum communication over repeater networks are discussed in~\cite{r1,r2,r3}.

To present the problem formally, we restrict attention, as in~\cite{Ugrinovskii2024,Ugrinovskii2024a}, to completely passive quantum systems. Accordingly, we use the annihilation-operator representation of their dynamics. Let \(u\) and \(w\) in Fig.~\ref{fig1} denote the vectors of annihilation operators associated with the engineered input field and the environment of the channel system, respectively. The vector of annihilation operators of the channel output field coupled with the equalizer is denoted $y$ in Fig. \ref{fig1}, and $z$ describes the annihilation operators associated with the equalizer's own environment. Also, the annihilation operators of the equalizer output field which are to be matched with $u$ are denoted $\hat{u}$ in the figure. We tacitly assume that the number of equalizer output modes is the same as the number of the driving modes of the channel, so $u$ and $\hat{u}$ consist of the same number of annihilation operators.

Drawing on the analogy with the classical case, consider the power spectrum density of the stationary component $e$ of the difference  ${u}-\hat{u}$ as a measure of the mean-square mismatch between $\hat{u}$ and $u$ \cite{valcdc,Ugrinovskii2024}:
\begin{equation}\label{eq1}
	P_e(i\omega)=\int_{-\infty}^{+\infty}\langle e(t)e(0)^\dagger\rangle e^{-i\omega t}dt;
\end{equation}
$\langle\cdot\rangle$ denotes the quantum expectation with respect to the underlying Gaussian state of the system. 
As in~\cite{Ugrinovskii2024,Ugrinovskii2024a}, the definition in \eqref{eq1} is stated for the annihilation-operator formulation of linear quantum systems. The coherent equalization problem consists in minimization of the power
spectrum density \eqref{eq1} over a class of coherent equalizers. Formally, it is to obtain an equalizer transfer function $H(s)$ which is realizable as a linear quantum system and attains or approximates closely the quantity
\begin{equation}\label{eq2}
\inf\sup_\omega\bar{\boldsymbol{\sigma}}(P_e(i\omega));
\end{equation}
$\bar{\boldsymbol{\sigma}}(\cdot)$ is the largest eigenvalue of a Hermitian matrix \cite{Ugrinovskii2024}. A quantum physical system that realizes such a transfer function is an optimal (respectively, near-optimal) coherent equalizer. The requirement for \emph{physical realizability} \cite{aline2011,hendra2017,sha2012} of the transfer function $H(s)$ is the essential constraint of the problem which sets it apart from classical filtering problems of similar kind.

It is worth mentioning that the coherent equalization problem \eqref{eq2} is substantially different from other coherent filtering problems considered in the recent quantum control literature and concerned with developments of the coherent quantum Kalman filter, the coherent Luenberger observer and coherent LQG control \cite{Miao2016,re2023,igor2013,Vuglar2014}. The differences can be seen in both the objectives of the filtering problem and the methodology applied to solve it. Indeed, Refs. \cite{Miao2016,igor2013,Vuglar2014}  are aimed at tracking internal modes of the underlying quantum system. The LQG performance cost in \cite{re2023} penalizes the classical internal state of a shaping filter and the output of a classical observer; the observer is then augmented with additional quantum inputs for physical realizability. In contrast, in the coherent equalization problem considered in this paper we seek to (near) optimally estimate the quantum system input in the mean-square sense using another quantum linear system. Also, the references \cite{Miao2016,re2023,igor2013,Vuglar2014} address the problem in the time domain. In contrast, the problem \eqref{eq2} is set in the frequency domain. The latter feature has proved instrumental in overcoming some of the technical difficulties of the time domain approach \cite{Ugrinovskii2024,igor2013}.

When both the channel and the equalizer are completely passive quantum systems, the problem \eqref{eq2} was shown in \cite{Ugrinovskii2024} to be reducible to a constrained optimization problem of $H_{\infty}$ type.
Specifically, for active channels or equalizers, the PSD should be replaced with an appropriate performance matrix formulation as described in~\cite{nurdin2009}. Then, under a certain $J$ spectral factorization assumption, this made it possible to characterize classes of near-optimal completely passive coherent equalizers in terms of a scattering transformation in the $H^{\infty}$ space of rational transfer functions. In \cite{Ugrinovskii2024a}, this approach was further advanced to show that the relaxation of the problem originally introduced in \cite{Ugrinovskii2024} is exact in the sense that the value in \eqref{eq2} can be approached arbitrarily closely by solving a certain auxiliary $H_{\infty}$ optimization problem, and then constructing a physically realizable $H(s)$ from its solution. This result was shown to hold under a certain additional condition; see \eqref{eq19} below. It guaranteed that any solution of the auxiliary problem, if it exists, could be used to construct a physically realizable $H(s)$ with a desired upper bound on \eqref{eq2}. The condition was shown to hold for quantum channels operating in a low signal-to-noise ratio environment, but it could fail when the environmental noise intensity was reduced.

In this paper we aim to circumvent the aforementioned sufficient condition. Here we observe that the auxiliary $H_{\infty}$ optimization problem which underpins the synthesis of coherent completely passive equalizers in \cite{Ugrinovskii2024a} is a special case of the so-called two-disk problem. The latter problem consists in finding a stabilizing $H_{\infty}$ controller which is itself stable and satisfies an $H_{\infty}$ norm condition \cite{zhou2001,norman1989numerical,ho1992}\footnote{In the general multi-disk problem, a feedback controller is sought to minimize the $H_{\infty}$ norm of a certain transfer function related to the closed loop system (e.g., the weighted sensitivity transfer function or the weighted complementary sensitivity transfer function), subject to constraints on the $H_{\infty}$ norm of other transfer functions \cite{norman1989numerical,ho1992}.
}, which is precisely the situation encountered here. This observation enables us to capitalize on the non-uniqueness of $H_{\infty}$ controllers/filters to select the one which satisfies two $H_{\infty}$ norm conditions simultaneously. This provides the freedom to select a solution to the auxiliary problem which ensures that the physical realizability requirements are satisfied.

Two-disk problems and more general strong $H_{\infty}$ stabilization problems were considered in \cite{zhou2001,Cao2000,cheng2008,ian2009,Zhu2019}. References \cite{Cao2000,cheng2008} used storage functions of specific form to ascertain the $H_{\infty}$ norm of the controller. To reduce the conservatism due to a specific choice of the storage functions,  \cite{Zhu2019,cheng2009} introduced additional slack variables. The design conditions resulting from this approach are generally expressed in terms of bilinear matrix inequalities (BMIs). Solving them requires additional relaxations or multiple alternating iterations, without a guarantee of convergence.

In this paper we address the two-disk $H_{\infty}$ problem related to the synthesis of completely passive coherent equalizers using the Youla type parameterization of all $H_{\infty}$ controllers \cite{zhou1996}. Effectively, the parameterization converts the underlying two-disk problem into another two-disk problem in which the central $H_{\infty}$ controller of the original problem defines the new plant, and the Youla parameter is treated as a feedback controller for it. Generally, such nested controller structures result in the controller order blow-out when dynamic parameters are used \cite{Cao2000,cheng2009}. To avoid this, we restrict attention to static parameters, however, we use non-minimal dynamic state space representations for them. This enables us to meet both $H_{\infty}$ norm conditions of the auxiliary problem at hand while the search for a suitable parameter becomes a convex feasibility problem for a set of linear matrix inequalities (LMIs) which can be solved efficiently using the existing methods of convex optimization \cite{lmi1994,nest1994}.

The main contribution of the paper is a computationally tractable procedure which allows one to construct a transfer function of a completely passive quantum equalizer for a given completely passive quantum communication system which guarantees a desired mean-square equalization performance. The proposed conditions for this are expressed in terms of feasibility of certain LMIs. We illustrate, via an example, that these conditions are considerably less
conservative than the conditions in \cite{Ugrinovskii2024a}. The gap between the near-optimal equalization performance ascertained via the proposed conditions and the optimal value \eqref{eq2} is also evaluated using a semidefinite program.

The paper is organized as follows. In the next section we state the coherent equalization problem formally and also present the background and some preliminary results from \cite{Ugrinovskii2024,Ugrinovskii2024a}. The conversion of the underlying coherent equalization problem into a two-disk $H_{\infty}$ problem is also presented in this section. Section \ref{sec3} is dedicated to the analysis of the general two-disk problem. The main results of the paper are presented in Section \ref{sec4} where the results of Section \ref{sec3} are applied to the coherent equalization problem under consideration in this paper. Section \ref{sec5} illustrates these results using a benchmark problem of equalization of a quantum cavity system from \cite{Ugrinovskii2024a}. Concluding remarks are given in Section \ref{sec6}.

\section{Preliminaries}\label{sec2}
\subsection{Open quantum linear systems}

We begin with reviewing the basics of open quantum systems; also see \cite{hendra2017,PhysRevA.81.023804,optics,Wiseman2009}. Consider a quantum channel system shown in Fig.~\ref{fig1}. The vectors of annihilation operators of the input field $u$ and $w$ consist of $n, n_{w}$ operators, respectively, $u=\operatorname{col}\left(u_{1}, \ldots, u_{n}\right), w=\operatorname{col}\left(w_{1}, \ldots, w_{n_{w}}\right)$; the symbol $\operatorname{col}(\cdot)$ denotes the column vector of operators defined in the underlying Hilbert space $\mathfrak{H}$. These operators satisfy the canonical commutation relations $\left[u_{j}(t), u_{k}^{*}\left(t^{\prime}\right)\right]=$ $\delta_{j k} \delta\left(t-t^{\prime}\right),\left[w_{j}(t), w_{k}^{*}\left(t^{\prime}\right)\right]=\delta_{j k} \delta\left(t-t^{\prime}\right)$; here $[\cdot, \cdot]$ is the commutator of two operators, $^*$ denotes the adjoint operator, $\delta(t)$ is the Dirac delta function, and $\delta_{j k}$ is the Kronecker symbol: $\delta_{j k}=1$ when $j=k$, otherwise $\delta_{j k}=0$. Also, $\left[u_{j}(t), w_{k}^{*}\left(t^{\prime}\right)\right]=0$. That is, the signal and environment operators commute. We also assume that the system is in a Gaussian thermal state $\rho$, and that the vectors of annihilation operators $u, w$ evolve as zero mean Gaussian quantum noise processes: $\langle u(t)\rangle=0,\langle w(t)\rangle=0$ where $\langle\cdot\rangle$ is the quantum expectation of the system in the state $\rho$ \cite{parth012}. Furthermore, the processes $u$ and $w$ are not correlated, $\left\langle u(t) w^{\dagger}\left(t^{\prime}\right)\right\rangle=0$.

Let $u^{\#}, w^{\#}$ denote the column vectors comprised of the adjoint operators of $u, w, u^{\#}=\operatorname{col}\left(u_{1}^{*}, \ldots, u_{n}^{*}\right)$, $w^{\#}=\operatorname{col}\left(w_{1}^{*}, \ldots, w_{n_{w}}^{*}\right)$. Also, define vectors of operators obtained by concatenating $u, u^{\#}$ and $w, w^{\#}$: $\breve{u}=\operatorname{col}\left(u, u^{\#}\right), \breve{w}=\operatorname{col}\left(w, w^{\#}\right)$. The autocorrelation matrices of the quantum processes $\breve{u}, \breve{w}$ are defined by
\begin{align}\label{eq3}
	& R_{\breve{u}}(t) \triangleq\left\langle\breve{u}(t) \breve{u}^{\dagger}(0)\right\rangle=\left[\begin{array}{cc}
		I+\Sigma_{u}^{T} & 0 \\
		0 & \Sigma_{u}
	\end{array}\right] \delta(t), \nonumber\\
	& R_{\breve{w}}(t) \triangleq\left\langle\breve{w}(t) \breve{w}^{\dagger}(0)\right\rangle=\left[\begin{array}{cc}
		I+\Sigma_{w}^{T} & 0 \\
		0 & \Sigma_{w}
	\end{array}\right] \delta(t) . 
\end{align}
The Hermitian \(n\times n\) and \(n_w\times n_w\) matrices \(\Sigma_u\) and \(\Sigma_w\) represent the signal and noise intensities in the quantum channel system shown in Fig.~\ref{fig1}, respectively~\cite{Ugrinovskii2024a}. Accordingly, they are positive semidefinite~\cite{Ugrinovskii2024a}.

The channel system consists of a collection of quantum harmonic oscillators interacting with the input quantum field. Its Hamiltonian and coupling with the input field involve only annihilation operators of the oscillator modes. 
Such systems are known as completely passive quantum systems~\cite{aline2011,hendra2017}, which include phase shifters, beam splitters, and optical cavities. These models are widely used for representing linear quantum channels in optical communication and quantum information processing~\cite{aline2011,PhysRevA.78.032323,5159845}. In the Heisenberg picture of quantum mechanics, the evolution of such systems can be described by the quantum stochastic differential equation in the Langevin form \cite{aline2011,hendra2017,PhysRevA.81.023804,optics,Wiseman2009,Gough2015}
\begin{align}\label{eq4}
\dot{\mathbf{a}}(t)&=A \mathbf{a}(t)+B_{1} u(t)+B_{2} w(t), \nonumber\\
	 y(t)&=C_{1} \mathbf{a}(t)+J_{11} u(t)+J_{12} w(t), \nonumber\\
	 d(t)&=C_{2} \mathbf{a}(t)+J_{21} u(t)+J_{22} w(t).
\end{align}
Here $A, B_{j}, C_{k}$ and $J_{k j}, j, k=1,2$, are complex matrices of dimensions $m_{s} \times m_{s}, m_{s} \times\left(n+n_{w}\right),\left(n_{y}+n_{d}\right) \times m_{s}$ and $\left(n_{y}+n_{d}\right) \times\left(n+n_{w}\right)$ respectively, and $\mathbf{a}$ is the vector of $m_{s}$ annihilation operators of the oscillator modes of the channel system. Also, $y, d$ are the $n_{y}$ component and $n_{d}$ component vectors of quantum noise processes corresponding to annihilation operators of the output field of the system; $y$ represents the part of the output field which is coupled with the equalizer system while $d$ represents the loss to the environment.

The transfer function $G(s)$ of the system \eqref{eq4} relates the bilateral Laplace transforms of $\operatorname{col}(u(t), w(t))$ and $\operatorname{col}(y(t), d(t))$ \cite{PhysRevA.81.023804,zhang2013}. It is partitioned to conform with the partition of the input and output processes,
\begin{equation*}
	G(s)=\left[\begin{array}{l}
		G_{11}(s)\quad G_{12}(s) \\
		G_{21}(s) \quad G_{22}(s)
	\end{array}\right],
\end{equation*}
where $G_{j k}(s)=C_{j}(s I-A)^{-1} B_{k}+J_{j k}, j, k=1,2$. Since the equation \eqref{eq4} and the corresponding transfer function $G(s)$ represent a physical quantum system, the transfer function matrix $G(s)$ is square (i.e., $n+n_{w}=n_{y}+n_{d}$) and paraunitary \cite{aline2011,sha2012,PhysRevA.81.023804,ball2013}:
\begin{equation}
	G(s)^{H} G(s)=G(s) G(s)^{H}=I_{n+n_{w}}. 
\end{equation}
Here and thereafter, $I_{k}$ is the $k \times k$ identity matrix, $G(s)^{H} \triangleq G\left(-s^{*}\right)^{\dagger}$, where the symbols ${ }^{*}$ and ${ }^{\dagger}$ denote the complex conjugate number and the complex conjugate transpose matrix, respectively.

It is assumed that the state space representation \eqref{eq4} of the transfer function $G(s)$ is minimal. Consequently the matrix $A$ is Hurwitz  \cite{Gough2015}, and $G(i \omega)$ is bounded at infinity and analytic on the entire closed imaginary axis \cite[Lemma~2]{youla1961}.

\subsection{Quantum coherent equalization problem}
This section describes the problem of coherent equalization introduced in \cite{valcdc,Ugrinovskii2024,Ugrinovskii2024a}. Also, some preliminary results from these references are reviewed.

As in \cite{Ugrinovskii2024,Ugrinovskii2024a}, we consider completely passive equalizers for the system \eqref{eq4}. The coherent filter is another open linear quantum system. It is coupled with the output field of the channel system \eqref{eq4} which carries information about the input field of the channel. It also interacts with its own environment. Mathematically, these interactions can also be described by a Langevin equation of the type \eqref{eq4} and the corresponding transfer function $H(s)$. The driving processes of the equalizer system are the vector $y$ of annihilation operators defined in the previous section, and the vector of annihilation operators $z=\operatorname{col}\left(z_{1}, \ldots, z_{n_{z}}\right)$ which represents the environment. The latter process evolves as a Gaussian zero-mean quantum noise process in the vacuum state and satisfies the canonical commutation relations $\left[z_{j}(t), z_{k}^{*}\left(t^{\prime}\right)\right]=\delta_{j k} \delta\left(t-t^{\prime}\right)$. That is, $\langle\breve{z}(t)\rangle=0$ where $\breve{z}=\operatorname{col}\left(z, z^{\#}\right)$, and the correlation function of the noise process $\breve{z}(t)$ is $\left\langle\breve{z}(t) \breve{z}^{\dagger}\left(t^{\prime}\right)\right\rangle=\begin{bmatrix}I_{n_{z}} & 0 \\ 0 & 0\end{bmatrix} \delta\left(t-t^{\prime}\right)$. It is assumed that $\breve{z}$ commutes with $\breve{u}$ and $\breve{w}$, and $\left\langle\breve{z}(t) \breve{u}^{\dagger}\left(t^{\prime}\right)\right\rangle=0$, $\left\langle\breve{z}(t) \breve{w}^{\dagger}\left(t^{\prime}\right)\right\rangle=0$. According to this description, the input field of the equalizer is associated with the $n_f$-dimensional vector of annihilation operators $\operatorname{col}(y,z)$, where $n_f=n_y+n_z$. The vector of the filter output processes has the same dimension $n_{f}$; it is partitioned into vectors $\hat{u}$ and $\hat{z}$ so that $\hat{u}$ has the same dimension $n$ as $u$.

For the transfer function $H(s)$ to represent a quantum physical system, it must have certain properties \cite{aline2011,sha2012,PhysRevA.81.023804}; they are summarized in the following definition.
\begin{definition}[\cite{Ugrinovskii2024a}, Definition 1]\label{def1}
	 An element $H(s)$ of the Hardy space $H_{\infty}$ is said to represent a completely passive physically realizable equalizer if $H(s)$ is a stable rational $n_{f} \times n_{f}$ transfer function, $n_{f}=n_{y}+n_{z} \geq n$, which is analytic in the right half-plane $\operatorname{Re}s>-\tau\;(\exists \tau>0)$ and is paraunitary,
	 \begin{equation}\label{eq6}
	 	H(s)^{H} H(s)=H(s) H(s)^{H}=I_{n_{f}}. 
	 \end{equation}
	\end{definition}

This definition formalizes the class of candidate equalizer transfer functions for the coherent equalization problem under consideration. Formally, the problem consists in finding a solution to the optimization problem \eqref{eq2} over the set of transfer functions described in Definition \ref{def1}.

It was shown in  \cite{valcdc,Ugrinovskii2024} that for any paraunitary $H(s)$,
\begin{equation}\label{eq7}
	P_e(i\omega)=\begin{bmatrix}H_{11}(i\omega)\; I\end{bmatrix}\Phi(i\omega)\begin{bmatrix}H_{11}(i\omega)^\dagger\\I\end{bmatrix}.
\end{equation}
Here $H_{11}(s)$ is an $n \times n_{y}$ transfer function which represents the top-left block of the equalizer transfer function $H(s)$ partitioned in accordance with the partitions of the equalizer input and output, $\operatorname{col}(y, z), \operatorname{col}(\hat{u}, \hat{z})$,
\begin{equation}
	H(s)=\begin{bmatrix}H_{11}(s)& H_{12}(s)\\[0.3em]H_{21}(s)& H_{22}(s)\end{bmatrix}.
\end{equation}
Also,
\begin{equation}\label{eq9}
\Phi(s) \triangleq\left[\begin{array}{cc}
	\Psi(s) & -G_{11}(s)\left(I_{n}+\Sigma_{u}^{T}\right)  \\
	-\left(I_{n}+\Sigma_{u}^{T}\right) G_{11}(s)^{H} & \Sigma_{u}^{T}+2 I_{n}
\end{array}\right],
\end{equation}
where
\begin{equation}\label{eq10}
	\Psi(s) \triangleq G_{11}(s) \Sigma_{u}^{T} G_{11}(s)^{H}+G_{12}(s) \Sigma_{w}^{T} G_{12}(s)^{H}.
\end{equation}

According to \eqref{eq7}, the PSD matrix $P_{e}$ is determined solely by the $(1,1)$ block of $H(s)$. This observation leads us to consider the following auxiliary problem.

\begin{definition}\label{def2}
	Given $\gamma>0$, the auxiliary problem is to obtain a proper rational $n\times n_y$ transfer function $H_{11}(s)$ with the following properties:
	\begin{itemize}
		\item[(a)] All poles of $H_{11}(s)$ are in the open left half-plane of	
		the complex plane, and $H_{11}(s)$ is analytic in a half-plane $\operatorname{Re}s>-\tau\;(\exists\tau>0)$;
		\item[(b)]
		\begin{equation}\label{eq11}
			H_{11}(i\omega)H_{11}(i\omega)^\dagger<I_n\quad\forall\omega\in\bar{\mathbf{R}},
		\end{equation}
		here $\bar{\mathbf{R}}$ is the closed real axis, $\bar{\mathbf{R}}=\mathbf{R}\cup\{\pm\infty\};$
		\item[(c)]
		\begin{equation}\label{eq12}
			P_e(i\omega)<\gamma^2I_n\quad\forall\omega\in\bar{\mathbf{R}}.
		\end{equation}
	\end{itemize}
The set of all transfer functions meeting the above requirements is denoted $\mathscr{H}_{11, \gamma}^{-}$.
\end{definition}

This auxiliary problem is central to the coherent equalizer synthesis methodology developed in \cite{Ugrinovskii2024a}. A method proposed in \cite{Ugrinovskii2024a} allows one to obtain an $H_{11} \in \mathscr{H}_{11, \gamma}^{-}$ and also construct from it a transfer function $H(s)$ which satisfies the requirements of Definition \ref{def1} and guarantees that
\begin{equation}
	\inf \sup _{\omega} \bar{\boldsymbol{\sigma}}\left(P_{e}(i \omega)\right)<\gamma^{2}. 
\end{equation}
Properties (a), (b) in Definition \ref{def2} are critical to be able to construct a physically realizable equalizer transfer function $H(s)$ from a solution of the auxiliary problem. The procedure for this was introduced in \cite{Ugrinovskii2024} and was revisited in \cite{Ugrinovskii2024a} under conditions (a), (b). According to \cite[Theorem 1]{Ugrinovskii2024a}, any $H_{11}$ which has properties (a), (b) in Definition \ref{def2} can be used as the $(1,1)$ block of a physically
realizable $H(s)$, and the remaining blocks $H_{12}, H_{21}, H_{22}$ of $H(s)$ can be obtained as follows. Let
\begin{equation}\label{eq14}
	\begin{aligned}
		Z_{1}(s)=&I_{n}-H_{11}(s) H_{11}(s)^{H},\\
		Z_{2}(s)=&I_{n_{y}}-H_{11}(s)^{H} H_{11}(s).
	\end{aligned}
\end{equation}
Then $H_{12}$ is taken to be a left spectral factor of $Z_{1}(s)$, i.e., a stable transfer function, analytic in the half-plane $\operatorname{Re}s >-\tau$ $(\exists \tau>0)$ and such that
\begin{equation}\label{eq15}
	Z_{1}(s)=H_{12}(s) H_{12}(s)^{H},
\end{equation}
and $H_{21}$ and $H_{22}$ are obtained as
\begin{align}
	& H_{21}(s)=U(s) \tilde{H}_{21}(s), \notag\\
	& H_{22}(s)=-U(s)(\tilde{H}_{21}^{-1}(s))^{H}H_{11}(s)^{H} H_{12}(s), \label{eq16}
\end{align}
where
\begin{itemize}
	\item $\tilde{H}_{21}(s)$ is a right spectral factor of $Z_{2}(s)$, i.e., a stable transfer function, analytic in the half-plane $\operatorname{Re} s>-\tau\;(\exists \tau>0)$ and such that
	\begin{equation}\label{eq17}
		Z_{2}(s)=\tilde{H}_{21}(s)^{H} \tilde{H}_{21}(s);
	\end{equation}
	\item $\tilde{H}_{21}^{-1}(s)$ denotes a right inverse of $\tilde{H}_{21}(s)$, i.e., a transfer function analytic in a right half-plane $\operatorname{Re}s>-\tau\;(\exists \tau>0)$ such that $\tilde{H}_{21}(s)\tilde{H}_{21}^{-1}(s)=I_r$, where $r$ is the normal rank of $Z_2(s)$ \cite{youla1961}; and
	\item $U(s)$ is a stable, analytic in the closed right half-plane, paraunitary $r \times r$ transfer function matrix, chosen to cancel unstable poles of $(\tilde{H}_{21}^{-1}(s))^{H} H_{11}(s)^{H}$ \cite{sha1990}.
\end{itemize}

\begin{rem}
	The above procedure requires weaker conditions on $H_{11}$ \cite{Ugrinovskii2024}. Namely, to construct a physically realizable $H(s)$ out of $H_{11}(s)$ with the properties described in (a) it suffices that $H_{11}(i \omega) H_{11}(i \omega)^{\dagger} \leq I_{n}$ for all $\omega$, and the normal rank of the matrices $Z_{1}(s)$ and $Z_{2}(s)$ does not change on the finite imaginary axis. It is shown in \cite{Ugrinovskii2024a} that \eqref{eq11} is sufficient for these conditions to hold.
\end{rem}

Performance of the coherent equalizer obtained using the above procedure has also been established in \cite{Ugrinovskii2024,Ugrinovskii2024a}. Specifically, Theorem 1 in \cite{Ugrinovskii2024a} shows that
\begin{equation}\label{eq18}
	\inf \sup _{\omega} \bar{\boldsymbol{\sigma}}\left(P_{e}(i \omega)\right) \leq \inf \left\{\gamma^{2}: \mathscr{H}_{11, \gamma}^{-} \neq \emptyset\right\}.
\end{equation}
Moreover, \cite[Theorem 2]{Ugrinovskii2024a}  shows that if there exists a $\theta>0$ such that
\begin{equation}\label{eq19}
\theta\left(\Phi(i \omega)-\gamma^{2}\left[\begin{array}{ll}
	0 & 0  \\
	0 & I_{n}
\end{array}\right]\right)-\left[\begin{array}{cc}
	I_{n_{y}} & 0 \\
	0 & -I_{n}
\end{array}\right]>0 \quad \forall \omega \in \bar{\mathbf{R}},
\end{equation}
then the inequality in \eqref{eq18} is the exact equality,
\begin{equation}\label{eq20}
	\inf\sup_\omega\bar{\boldsymbol{\sigma}}(P_e(i\omega))=\inf\{\gamma^2:\mathscr{H}_{11,\gamma}^-\neq\emptyset\}.
\end{equation}
As a result, the optimal equalization error \eqref{eq2} can be approximated with an arbitrary desired precision by minimizing over all $\gamma$ for which the auxiliary problem in Definition \ref{def2} has a nonempty solution set, i.e., by solving the optimization problem on the right hand side of \eqref{eq20} and \eqref{eq18},
\begin{equation}\label{eq21}
	\gamma_{\circ}^{\prime \prime} \triangleq \inf \left\{\gamma>0: \mathscr{H}_{11, \gamma}^{-} \neq \emptyset\right\}.
\end{equation}
Also, a coherent equalizer transfer function $H(s)$ which delivers a near-optimal mean-square equalization performance can be constructed from a solution of the auxiliary problem involving the near-optimal $\gamma$ of the problem \eqref{eq21}.

As noted, these results are contingent on the satisfaction of condition \eqref{eq19}. Under this condition any transfer function $H_{11}$ which has properties (a) and (c) in Definition \ref{def2} is also a contraction in the sense of \eqref{eq11}, i.e., property (b) also holds. Technically, \eqref{eq19} renders condition \eqref{eq11} inactive as a constraint of the optimization problem \eqref{eq21}. As a result, this problem reduces to a standard optimal $H_{\infty}$ filtering problem. Such problem is considerably easier to solve than the original optimization problem \eqref{eq21}, however this solution path requires \eqref{eq19} to hold. The latter condition turns out to be conservative \cite{Ugrinovskii2024a}.

In this paper we develop a different approach to finding a solution to the auxiliary problem in Definition \ref{def2} and the related optimization problem \eqref{eq21}, which does not rely on condition \eqref{eq19}. In this approach, condition \eqref{eq11} is incorporated in the design algorithm, in contrast to rendering it inactive as was done in \cite{Ugrinovskii2024a}. As a result, the auxiliary problem becomes a type of the two-disk $H_{\infty}$ control problem.

\subsection{A two-disk formulation of the auxiliary problem}
The results in this paper require the following technical assumption.
\begin{assumption} [see Assumption 1 in \cite{Ugrinovskii2024a}]\label{ass1}
 There exists a constant $\lambda \geq 0$ such that the $\left(n_{y}+n\right) \times\left(n_{y}+n\right)$ rational matrix transfer function
	\begin{equation}
		\Phi_{\lambda}(s)=\Phi(s)+\left[\begin{array}{cc}
			0 & 0 \\
			0 & \lambda^{2} I_{n}
		\end{array}\right]
	\end{equation}
	admits a spectral factorization
	\begin{equation}\label{eq23}
		\Phi_{\lambda}(s)=\Upsilon_{\lambda}(s) \Upsilon_{\lambda}(s)^{H}, 
	\end{equation}
	where a $\left(n_{y}+n\right) \times p$ rational transfer matrix $\Upsilon_{\lambda}(s)$ has all its poles in the left half-plane $\operatorname{Re} s<-\tau$ and is analytic in $\operatorname{Re} s>-\tau\;(\exists \tau>0)$.
\end{assumption}
\begin{rem}
		As shown in~\cite[Lemma~1 and Remark~4]{Ugrinovskii2024a}, such a \(\lambda\) always exists when \(\Sigma_u>0\). However, for a vacuum input field one has \(\Sigma_u=0\), in which case the above existence result for Assumption~\ref{ass1} is no longer guaranteed.
\end{rem}

Consider a minimal state space realization of the spectral factor $\Upsilon_{\lambda}(s)$ in \eqref{eq23}:
\begin{equation}\label{eq24}
\Upsilon_{\lambda}(s) \sim\left[\begin{array}{c|c}
	A_{\lambda} & B_{\lambda}  \\
	\hline C_{1, \lambda} & D_{1, \lambda} \\
	C_{2, \lambda} & D_{2, \lambda}
\end{array}\right].
\end{equation}
The complex matrices $A_{\lambda}, B_{\lambda}, C_{1, \lambda}, C_{2, \lambda}, D_{1, \lambda}, D_{2, \lambda}$ have dimensions $m \times m, m \times p, n_{y} \times m, n \times m, n_{y} \times p, n \times p$, respectively, where $m$ is the McMillan degree of $\Upsilon_{\lambda}(s)$. Note that the output of $\Upsilon_{\lambda}$ is partitioned into vectors of dimensions $n_{y}$ and $n$ respectively; that is, $\Upsilon_{\lambda}=\left[\begin{array}{c}\Upsilon_{1, \lambda} \\ \Upsilon_{2, \lambda}\end{array}\right]$ where $\Upsilon_{j, \lambda}=C_{j, \lambda}\left(s I_{m}-A_{\lambda}\right)^{-1} B_{\lambda}+D_{j, \lambda}, j=1,2$. Also, according to Assumption \ref{ass1}, the matrix $A_{\lambda}$ is Hurwitz.

Introduce the following transfer functions and a constant $\bar{\gamma}$:
\begin{align}
	& \bar{H}_{11}(s)=H_{11}\left(s^{*}\right)^{\dagger}, \quad \bar{\Upsilon}_{\lambda}(s)=\Upsilon_{\lambda}\left(s^{*}\right)^{\dagger}, \nonumber\\
	& \bar{\gamma}=\left(\gamma^{2}+\lambda^{2}\right)^{1 / 2}.
\end{align}
Note that $\bar{\Upsilon}_{\lambda}(s)$ is partitioned as $\left[\bar{\Upsilon}_{1, \lambda}(s) \quad \bar{\Upsilon}_{2, \lambda}(s)\right]$ where $\bar{\Upsilon}_{j, \lambda}(s)=\Upsilon_{j, \lambda}\left(s^{*}\right)^{\dagger}, j=1,2$.

Our development relies on the following result from \cite{Ugrinovskii2024a}.
\begin{lemma}[Lemma 2, \cite{Ugrinovskii2024a}]\label{lem1}
 A stable proper transfer function $H_{11}(s)$ which is analytic in a half-plane $\operatorname{Re}s >-\tau\;(\exists \tau>0)$ satisfies \eqref{eq12} if and only if
	\begin{equation}\label{eq26}
		T_{\lambda}(i \omega)^{\dagger} T_{\lambda}(i \omega)<\bar{\gamma}^{2} I_{n} \quad \forall \omega \in \bar{\mathbf{R}}, 
	\end{equation}
	where
\begin{equation}\label{eq27}
	T_{\lambda}(s) \triangleq \bar{\Upsilon}_{\lambda}(s)\left[\begin{array}{c}
	\bar{H}_{11}(s)  \\
	I_{n}
\end{array}\right].
\end{equation}
\end{lemma}

\begin{figure}
	\centering
	\includegraphics[width=1\columnwidth]{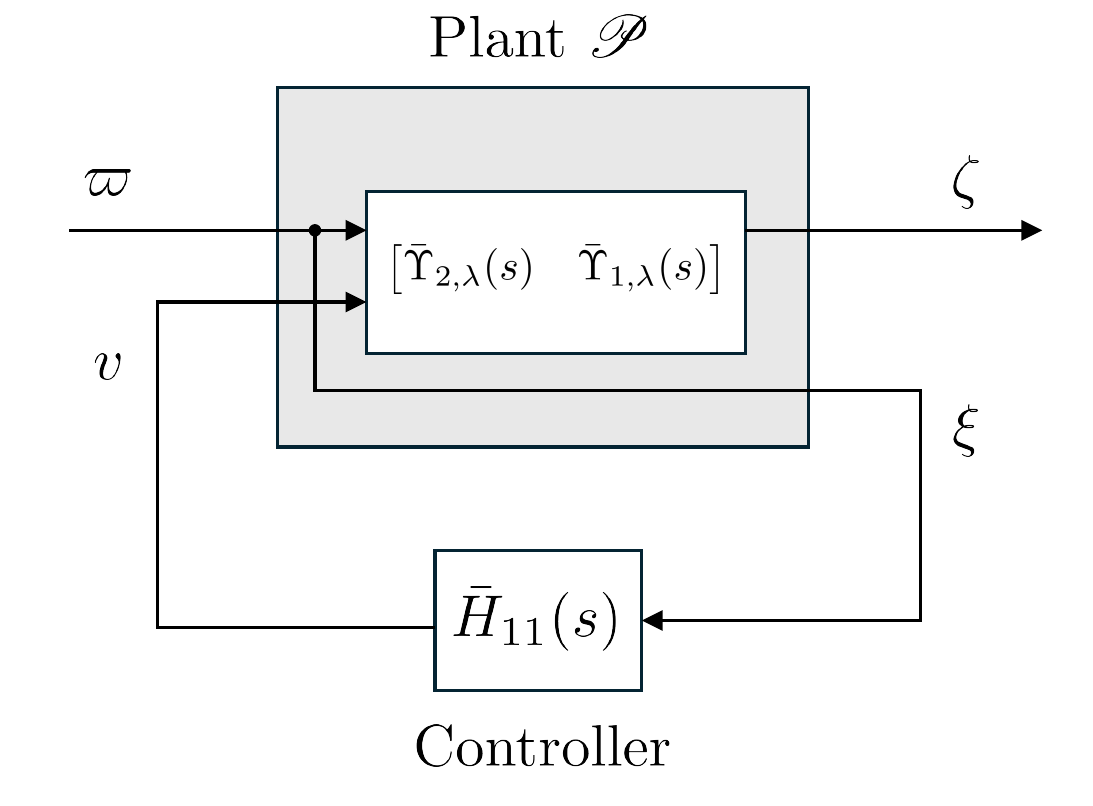}
{\caption{The $H_\infty$ control setting for the auxiliary problem.}\label{fig2}}
\end{figure}

It is straightforward to show that the transfer function $T_{\lambda}(s)$ in \eqref{eq27} is the linear fractional transformation \footnote{Given a four-block plant $\mathscr{P}:\left[\begin{array}{l}\zeta \\ \xi\end{array}\right]=\left[\begin{array}{ll}\mathscr{P}_{11} & \mathscr{P}_{12} \\ \mathscr{P}_{21} & \mathscr{P}_{22}\end{array}\right]\left[\begin{array}{l}\varpi \\ v\end{array}\right]$ and a feedback controller $v=\mathscr{C} \xi$, the linear fractional transformation $\mathscr{F}_{l}(\mathscr{P}, \mathscr{C})$ defines the $\varpi \rightarrow \zeta$ transfer function of the closed loop system obtained by interconnecting $\mathscr{P}$ and $\mathscr{C}$ \cite{zhou1996,Safonov1989}:
	\begin{equation*}
		\mathscr{F}_{l}(\mathscr{P}, \mathscr{C})=\mathscr{P}_{11}+\mathscr{P}_{12} \mathscr{C}\left(I-\mathscr{P}_{22} \mathscr{C}\right)^{-1} \mathscr{P}_{21}.
	\end{equation*}
} involving the plant $\mathscr{P}$ shown in Fig. \ref{fig2} and the feedback controller $\bar H_{11}$:
\begin{equation*}
T_{\lambda}(s)=\mathscr{F}_{l}\left(\mathscr{P}, \bar{H}_{11}\right).
\end{equation*}

The plant $\mathscr{P}$ consists of the system with the transfer function $\left[\bar{\Upsilon}_{2, \lambda}(s) \quad \bar{\Upsilon}_{1, \lambda}(s)\right]$ augmented with the disturbance feedthrough output $\xi=\varpi$, which serves as a `measurement output' of the plant $\mathscr{P}$. That is, $\mathscr{P}$ has the following four block representation
\begin{equation*}
\left[\begin{array}{l}
	\zeta \\
	\xi
\end{array}\right]=\left[\begin{array}{cc}
	\bar{\Upsilon}_{2, \lambda}(s) & \bar{\Upsilon}_{1, \lambda}(s) \\
	I_{n} & 0
\end{array}\right]\left[\begin{array}{l}
	\varpi \\
	v
\end{array}\right],
\end{equation*}
where $\varpi \in \mathbf{C}^{n}, v \in \mathbf{C}^{n_{y}}, \zeta \in \mathbf{C}^{p}$, and $\xi \in \mathbf{C}^{n}$. The output $\xi$ is used as the input to the feedback controller $\bar{H}_{11}$. The state-space representation of the plant $\mathscr{P}$ is obtained from \eqref{eq24}:
\begin{equation}\label{eq281}
	\begin{aligned}
		\dot{x}=&A_{\lambda}^{\dagger} x+C_{2, \lambda}^{\dagger} \varpi+C_{1, \lambda}^{\dagger} v,\\
		\zeta=&B_{\lambda}^{\dagger} x+D_{2, \lambda}^{\dagger} \varpi+D_{1, \lambda}^{\dagger} v,\\
		\xi=&\varpi.
	\end{aligned}
\end{equation}
Here $x \in \mathbf{C}^{m}$. It will be convenient to use the standard\\
notation for $\mathscr{P}$,
\begin{equation}\label{eq29}
	 \mathscr{P} \sim\left[\begin{array}{c|cc}
		\bar{A} & \bar{B}_{1} & \bar{B}_{2} \\
		\hline \bar{C}_{1} & \bar{D}_{11} & \bar{D}_{12}  \\
		\bar{C}_{2} & \bar{D}_{21} & \bar{D}_{22}
	\end{array}\right] , \\
\end{equation}
\begin{equation}\label{eq30}
	\begin{aligned}
		&\bar A=A_{\lambda}^{\dagger},\quad\bar C_{1}=B_{\lambda}^{\dagger},\quad\bar C_{2}=0, \\
		&\bar{B}_{1}=C_{2,\lambda}^{\dagger},\quad\bar{B}_{2}=C_{1,\lambda}^{\dagger}, \\
		&\bar D=\begin{bmatrix}\bar D_{11}& \bar D_{12}\\[0.3em]\bar D_{21}& \bar D_{22}\end{bmatrix}=\begin{bmatrix}D_{2,\lambda}^\dagger &D_{1,\lambda}^\dagger\\I_n&0\end{bmatrix}.
	\end{aligned}
\end{equation}

These observations and Lemma \ref{lem1} transform the auxiliary problem of finding $H_{11} \in \mathscr{H}_{11, \gamma}^{-}$ into an equivalent problem of feedback control design. In this problem, a rational feedback controller $\bar{H}_{11}(s)$ is to be found which guarantees that the closed loop system $\mathscr{F}_{l}\left(\mathscr{P}, \bar H_{11}\right)$ is internally stable and has the $H_{\infty}$ norm less than $\bar{\gamma}=\left(\gamma^{2}+\lambda^{2}\right)^{1 / 2}$, and also the controller transfer function $\bar{H}_{11}(s)$ is required to be a stable and strictly contractive element of $H_{\infty}$:
\begin{equation}\label{eq31}
	\left\|\mathscr{F}_{l}\left(\mathscr{P}, \bar{H}_{11}\right)\right\|_{\infty}<\bar{\gamma}, \quad \text { and } \quad\left\|\bar{H}_{11}\right\|_{\infty}<1.
\end{equation}
This problem is a special case of the two-disk $H_{\infty}$ control problem.

\section{An LMI synthesis of two-disk $H_{\infty}$ controllers}\label{sec3}
The two-disk problem is concerned with finding an internally stabilizing $n_{y} \times n$ controller $\mathscr{C}$ for a plant $\tilde{\mathscr{P}}$ such that $\mathscr{C} \in H_{\infty}$ and
\begin{align}
	& \left\|\mathscr{F}_{l}(\tilde{\mathscr{P}}, \mathscr{C})\right\|_{\infty}<\tilde{\gamma}, \label{eq32} \\
	& \left\|V \mathscr{C}(s) W^{\dagger}-J\right\|_{\infty}<1. \label{eq33}
\end{align}
In these inequalities, $\tilde{\gamma}>0$ is a given constant. 
Here, \( V \), \( W \), and \( J \) are complex matrices of dimensions \( n_{y} \times n_{y} \), \( n \times n \), and \( n_{y} \times n \), respectively, with \( V \) and \( W \) assumed nonsingular. Their specific forms in the context of coherent quantum equalization will be given in \eqref{eq57}. Condition \eqref{eq33} generalizes the requirement $\|\mathscr{C}\|_{\infty}<1$ of the two-disk problem \eqref{eq31}, it arises after loop-shifting transformations \cite{zhou1996,Safonov1989} are applied to the plant $\mathscr{P}$ in order to transform it to the plant $\tilde{\mathscr{P}}$ that has a so-called standard form,
\begin{equation}\label{eq34}
	\begin{aligned}
		\dot{x}=&\tilde{A} x+\tilde{B}_{1} \varpi+\tilde{B}_{2} v,\\
		\zeta=&\tilde{C}_{1} x+\tilde{D}_{12} v,\\
		\xi=&\tilde{C}_{2} x+\tilde{D}_{21} \varpi.
	\end{aligned}
\end{equation}
In the synthesis of coherent equalizers presented in the next section, and under Assumption~\ref{ass2}, \eqref{eq281} is reformulated into the form of \eqref{eq34} using \eqref{eq55}.
Therefore in the remainder of this section we assume that the plant $\tilde{\mathscr{P}}$ has the form \eqref{eq34} and also satisfies the following assumptions \cite{zhou1996}:

(A1) $(\tilde{A}, \tilde{B}_{2})$ is stabilizable and $(\tilde{C}_{2}, \tilde{A})$ is detectable;\\
(A2) $\tilde{D}_{12}=\left[\begin{array}{c}0 \\ I_{n_{y}}\end{array}\right], \tilde{D}_{21}=\left[\begin{array}{ll}0 & I_{n}\end{array}\right]$;

(A3) $\left[\begin{array}{cc}\tilde{A}-s I_{m} & \tilde{B}_{2} \\ \tilde{C}_{1} & \tilde{D}_{12}\end{array}\right]$ has full column rank for all $s, \operatorname{Re}s \geq 0$;

(A4) $\left[\begin{array}{cc}\tilde{A}-s I_{m} & \tilde{B}_{1} \\ \tilde{C}_{2} & \tilde{D}_{21}\end{array}\right]$  has full row rank for all $s, \operatorname{Re} s \geq 0$.

Assumption (A2) is not critical and is easily satisfied via a series of loop shifting transformations \cite{Safonov1989}. On the other hand, Assumptions (A1), (A3) and (A4) are necessary for the existence of the solution of the $H_{\infty}$ synthesis problem encapsulated in \eqref{eq32}.

The technical foundation of our solution to the two-disk problem \eqref{eq32}, \eqref{eq33} involves the complex versions of the bounded real lemma and the elimination/projection lemma \cite{dull2013}. These techniques have been applied previously to two-disk problems; e.g., see \cite{Zhu2019}. Our approach differs in that we apply these techniques to the parameterization of all $H_{\infty}$ controllers, instead of applying them to the closed loop system comprised of the plant $\tilde{\mathscr{P}}$ and a dynamic controller $\mathscr{C}$. First, we recall that every solution to the $H_{\infty}$ control problem \eqref{eq32} can be expressed as the linear fractional transformation involving the central controller $K_{c}$ of the $H_{\infty}$ control problem \eqref{eq32} and an $n_{y} \times n$ matrix transfer function $\mathscr{Q}(s) \in H_{\infty}$ such that $\|\mathscr{Q}\|_{\infty} \leq \tilde{\gamma}$ \cite{zhou1996}. Therefore, expressing $\mathscr{C}(s)$ as
\begin{equation}\label{eq35}
	\mathscr{C}(s)=\mathscr{F}_{l}\left(K_{c}, \mathscr{Q}\right) 
\end{equation}
ensures that the first $H_{\infty}$ norm condition \eqref{eq32} is satisfied and also provides the freedom to choose $\mathscr{Q}$ of an arbitrary order or impose additional constraints on $\mathscr{Q}$. For instance, we will be able to select a static $\mathscr{Q}$. This will give us an easy way to ensure that $\|\mathscr{Q}\|_{\infty} \leq \tilde{\gamma}$, while the order of the controller $\mathscr{C}$ is kept the same as the order of the central controller $K_{c}$.

Recall that the state space model of the central controller $K_{c}$
\begin{equation}\label{eq36}
K_{c} \sim\left[\begin{array}{c|cc}
	A_{K_{c}} & B_{K_{c}, 1} & B_{K_{c}, 2}  \\
	\hline C_{K_{c}, 1} & 0 & I_{n_{y}} \\
	C_{K_{c}, 2} & I_{n} & 0
\end{array}\right],
\end{equation}
is obtained from the stabilizing Hermitian positive semidefinite solutions to certain Riccati equations. The formulas for the coefficients of this state space model can be found in \cite[p.444]{zhou1996}.

We are now in a position to present a result about the two-disk problem \eqref{eq32}, \eqref{eq33}. Introduce a state space realization of the transfer function $\mathscr{Q}$ and the matrix of its parameters,
\begin{equation}
\mathscr{Q} \sim\left[\begin{array}{c|c}A_{Q} & B_{Q} \\ \hline C_{Q} & D_{Q}\end{array}\right], \quad Q=\left[\begin{array}{cc}A_{Q} & B_{Q} \\ C_{Q} & D_{Q}\end{array}\right] ;
\end{equation}
$A_{Q}, B_{Q}, C_{Q}, D_{Q}$ are complex matrices of dimensions $m \times m, m \times n, n_{y} \times m, n_{y} \times n$. Next, define $N_{o}, N_{c}$ to be full rank matrices which span the null spaces of $[C_{K_{c}, 2} \quad\ W^{\dagger}]$ and $[B_{K_{c}, 2}^{\dagger}\quad V^{\dagger}]$, respectively:
\begin{equation*}
	\operatorname{Im} N_{o}=\operatorname{Ker}\left[\begin{array}{ll}C_{K_{c}, 2} & W^{\dagger}\end{array}\right], \; \operatorname{Im} N_{c}=\operatorname{Ker}\left[\begin{array}{ll}B_{K_{c}, 2}^{\dagger} & V^{\dagger}\end{array}\right].
\end{equation*}

Also, introduce matrix inequalities in the variables $X_{1}=$ $X_{1}^{\dagger} \in \mathbf{C}^{m \times m}, Y_{1}=Y_{1}^{\dagger} \in \mathbf{C}^{m \times m}$:
\begin{equation}\label{eq38}
X_{1}>0, \quad Y_{1}>0,
\end{equation}
\begin{equation}
	\begin{aligned}\begin{bmatrix}N_o^\dagger&0\\0&I_{n_y}\end{bmatrix}\begin{bmatrix}A_{K_c}^\dagger X_1+X_1A_{K_c}&X_1B_{K_c,1}W^\dagger&C_{K_c,1}^\dagger V^\dagger\\WB_{K_c,1}^\dagger X_1&-I_n&-J^\dagger\\VC_{K_c,1}&-J&-I_{n_y}\end{bmatrix}\\
		\times\begin{bmatrix}N_o&0\\0&I_{n_y}\end{bmatrix}<0,\end{aligned}
\end{equation}
\begin{equation}
	\begin{aligned}\begin{bmatrix}N_c^\dagger&0\\0&I_n\end{bmatrix}\begin{bmatrix}A_{K_c}Y_1+Y_1A_{K_c}^\dagger&Y_1C_{K_c,1}^\dagger V^\dagger&B_{K_c,1}W^\dagger\\VC_{K_c,1}Y_1&-I_{n_y}&-J\\WB_{K_c,1}^\dagger&-J^\dagger&-I_n\end{bmatrix}\\
		\times\begin{bmatrix}N_c&0\\0&I_n\end{bmatrix}<0,\end{aligned}
\end{equation}
\begin{equation}\label{eq41}
\left[\begin{array}{ll}X_{1} & I_{m} \\ I_{m} & Y_{1}\end{array}\right] \geq 0,
\end{equation}
and the matrix
\begin{equation}\label{eq42}
\hat{X}=\left[\begin{array}{cc}
	X_{1} & X_{2}  \\
	X_{2}^{\dagger} & I_{m}
\end{array}\right],
\end{equation}
where an $m \times m$ matrix $X_{2}$ is chosen so that $X_{2} X_{2}^{\dagger}=$ $X_{1}-Y_{1}^{-1}$. Finally, define the matrices
\begin{align}
	\hat{A} & =\left[\begin{array}{cc}
		A_{K_{c}} & 0 \\
		0 & 0
	\end{array}\right], \quad \hat{B}=\left[\begin{array}{c}
		B_{K_{c}, 1} W^{\dagger} \\
		0
	\end{array}\right], \nonumber\\
	\hat{C} & =\left[\begin{array}{ll}
		V C_{K_{c}, 1} & 0
	\end{array}\right], \quad \hat{D}=-J, \nonumber\\
	\underline{B} & =\left[\begin{array}{cc}
		0 & B_{K_{c}, 2} \\
		I & 0
	\end{array}\right], \quad \underline{C}=\left[\begin{array}{cc}
		0 & I \\
		C_{K_{c}, 2} & 0
	\end{array}\right], \nonumber\\
	\underline{D}_{12} & =\left[\begin{array}{cc}
		0 & V
	\end{array}\right], \quad \underline{D}_{21}=\left[\begin{array}{c}
		0 \\
		W^{\dagger}
	\end{array}\right],  \\
	\Sigma_{\hat{X}} & =\left[\begin{array}{ccc}
		\hat{A}^{\dagger} \hat{X}+\hat{X} \hat{A} & \hat{X} \hat{B} & \hat{C}^{\dagger} \\
		\hat{B}^{\dagger} \hat{X} & -I_{n} & \hat{D}^{\dagger} \\
		\hat{C} & \hat{D} & -I_{n_{y}}
	\end{array}\right], \nonumber\\
	\Lambda_{\hat{X}} & =\left[\begin{array}{lll}
		\underline{B}^{\dagger} \hat{X} & 0 & \underline{D}_{12}^{\dagger}
	\end{array}\right], \quad \Pi=\left[\begin{array}{lll}
		\underline{C} & \underline{D}_{21} & 0
	\end{array}\right].\nonumber
\end{align}

\begin{theorem}\label{th1}
	Suppose conditions (A1)-(A4) are satisfied and the inequalities \eqref{eq38}-\eqref{eq41} are feasible. Let $\hat{X}$ be the matrix defined in \eqref{eq42}. If the linear matrix inequalities
	\begin{align}
		& \Sigma_{\hat{X}}+\Lambda_{\hat{X}}^{\dagger} Q \Pi+\Pi^{\dagger} Q^{\dagger} \Lambda_{\hat{X}}<0, \label{eq44}  \\
		& {\left[\begin{array}{cc}
				\tilde{\gamma}^{2} I_{n_{y}} & D_{Q} \\
				D_{Q}^{\dagger} & I_{n}
			\end{array}\right]>0} \label{eq45},
	\end{align}
	in which the variable $Q$ is restricted to have the form
	\begin{equation}\label{eq46}
		Q=\left[\begin{array}{cc}
		A_{Q} & 0  \\
		C_{Q} & D_{Q}
	\end{array}\right],
	\end{equation}
	have a feasible solution, then the controller	
	\begin{align}
		& \mathscr{C} \sim\left[\begin{array}{c|c}A_{\mathscr{C}} & B_{\mathscr{C}} \\ \hline C_{\mathscr{C}} & D_{\mathscr{C}}\end{array}\right],\\
		& A_{\mathscr{C}}=A_{K_{c}}+B_{K_{c}, 2} D_{Q} C_{K_{c}, 2},  \nonumber\\
		& B_{\mathscr{C}}=B_{K_{c}, 1}+B_{K_{c}, 2} D_{Q},  \nonumber\\
		& C_{\mathscr{C}}=C_{K_{c}, 1}+D_{Q} C_{K_{c}, 2},  \nonumber\\
		& D_{\mathscr{C}}=D_{Q}, \label{eq47}
	\end{align}
	solves the two-disk problem \eqref{eq32}, \eqref{eq33}.
\end{theorem}
\begin{pf}
 Taking \eqref{eq35} into \eqref{eq33} converts the original two-disk problem \eqref{eq32}, \eqref{eq33} into another two-disk problem
\begin{equation}\label{eq48}
	\left\|V \mathscr{F}_{l}\left(K_{c}, \mathscr{Q}\right) W^{\dagger}-J\right\|_{\infty}<1, \quad\|\mathscr{Q}\|_{\infty}<\tilde{\gamma}. 
\end{equation}
In this problem, $\mathscr{K}=V \mathscr{F}_{l}\left(K_{c}, \mathscr{Q}\right) W^{\dagger}-J$ plays the role of the closed loop plant, while $\mathscr{Q} \in H_{\infty}$ plays the role of a stabilizing $H_{\infty}$ controller. Our proof is focused on this problem, since the norm condition \eqref{eq32} will be implied if a solution to \eqref{eq48} exists. Also, \eqref{eq33} is the same as the first condition \eqref{eq48} .

According to \cite[Chapter 7]{dull2013}, the inequality \eqref{eq44} implies that the controller
\begin{align}
	& \mathscr{C} \sim\left[\begin{array}{l|l}
		A_{\mathscr{C}} & B_{\mathscr{C}} \\
		\hline C_{\mathscr{C}} & D_{\mathscr{C}}
	\end{array}\right], \nonumber\\
	& A_{\mathscr{C}}=\left[\begin{array}{cc}
		A_{K_{c}}+B_{K_{c}, 2} D_{Q} C_{K_{c}, 2} & B_{K_{c}, 2} C_{Q} \\
		0_{m \times m} & A_{Q}
	\end{array}\right], \nonumber\\
	& B_{\mathscr{C}}=\left[\begin{array}{c}
		B_{K_{c}, 1}+B_{K_{c}, 2} D_{Q} \\
		0_{m \times n}
	\end{array}\right], \nonumber\\
	& C_{\mathscr{C}}=\left[\begin{array}{ll}
		C_{K_{c}, 1}+D_{Q} C_{K_{c}, 2} & C_{Q}
	\end{array}\right], \quad D_{\mathscr{C}}=D_{Q} \label{eq49} 
\end{align}
renders the first disk condition \eqref{eq48} true and also guarantees that the matrix $A_{\mathscr{C}}$ is Hurwitz. That is, $A_{K_{c}}+$ $B_{K_{c}, 2} D_{Q} C_{K_{c}, 2}$ and $A_{Q}$ are both Hurwitz. As a result, both $\mathscr{Q}$ and $\mathscr{C}$ lie in $H_{\infty}$. Furthermore, the state space realization of $\mathscr{C}$ in \eqref{eq49} is not controllable since the controllability matrix of $\left(A_{\mathscr{C}}, B_{\mathscr{C}}\right)$,
\begin{equation*}
\left[\begin{array}{cc}
	\Gamma_{c} & \left(A_{K_{c}}+B_{K_{c}, 2} D_{Q} C_{K_{c}, 2}\right)^{m} \Gamma_{c} \\
	0 & 0
\end{array}\right],
\end{equation*}
has $m$ zero rows; here $\Gamma_{c}$ denotes the controllability matrix of the pair $\left(A_{K_{c}}\!+\!B_{K_{c}, 2} D_{Q} C_{K_{c}, 2}, B_{K_{c}, 1}\!+\!B_{K_{c}, 2} D_{Q}\right)$. Reducing $\mathscr{C}$ by removing uncontrollable states yields the state space model \eqref{eq47} for $\mathscr{C}$. Clearly, the transfer function $\mathscr{K}$ is not affected by this, so the first disk condition \eqref{eq48} still holds. As noted, this condition is the same as \eqref{eq33}.

Moreover, since $B_{Q}=0$, the minimal realization of $\mathscr{Q}$ is $\mathscr{Q}=D_{Q}$. Condition \eqref{eq45} yields $\|\mathscr{Q}\|_{\infty}<\tilde{\gamma}$. Thus, the controller $\mathscr{C}$ solves the $H_{\infty}$ control problem for the plant $\tilde{\mathscr{P}}$, i.e., the disk condition \eqref{eq32} is also satisfied.   \hfill $\Box$
\end{pf}
\begin{rem}
	The linear matrix inequality \eqref{eq44} involves two slack variables $A_{Q}$ and $C_{Q}$ of dimensions $m \times m, n_{y} \times m$. By introducing these variables, we are able to express the conditions of Theorem \ref{th1} in terms of linear matrix inequalities. The slack variables can be reduced to have dimensions $m_{Q} \times m_{Q}, n_{y} \times m_{Q}, m_{Q}<m$, or even eliminated altogether at the expense of adding the rank condition
	\begin{equation}
			\operatorname{rank}\left[\begin{array}{cc}
			X_{1} & I  \\
			I & Y_{1}
		\end{array}\right] \leq m+m_{Q}.
	\end{equation}
	However, the convexity of the conditions of the theorem will be lost, as a result. Therefore, we set \( B_Q = 0 \), which increases computational complexity compared to a fully Hermitian LMI but avoids introducing non-convex rank constraints, thereby preserving tractability.
\end{rem}
\begin{rem}
	Feasibility of the matrix inequalities \eqref{eq38}-\eqref{eq41} is both sufficient and necessary for the existence of a controller $\mathscr{C}$ which satisfies \eqref{eq32}. Yet, Theorem \ref{th1} gives only a sufficient condition for \eqref{eq33}. The gap is due to the particular form of the parameter $\mathscr{Q}$ whose input matrix $B_{Q}$ is set to $0$ to facilitate the property $\|\mathscr{Q}\|_{\infty}<\tilde{\gamma}$ and also to ensure that the order of the controller matches the order of the central controller. It may be possible to tighten the gap, e.g., by introducing different slack variables or using a dynamic $\mathscr{Q}$ or by embedding additional feedback layers in the controller. However, these tighter conditions will be achieved at a cost. The additional layers of feedback will increase the dynamic order of the controller. When we employ such a solution to synthesize a coherent equalizer, such an equalizer will be more complex to implement as a result. Other approaches to reducing conservatism via introducing slack variables were shown to lead to nonconvex BMI conditions for the existence of the controller \cite{Zhu2019,cheng2009}, which were considerably more difficult to solve.
	\end{rem}
\begin{rem}
	Instead of letting $B_{Q}=0$, we can let $C_{Q}=0$ in \eqref{eq46} and let $B_{Q}$ be a slack variable. The proof of the theorem will not change, except the controller \eqref{eq49} will become unobservable rather than uncontrollable. However, our simulations reveal that the resulting conditions tend to be more conservative.
	\end{rem}

\section{Synthesis of coherent equalizers}\label{sec4}
\subsection{A solution to the auxiliary two-disk problem \eqref{eq31}}
To apply Theorem \ref{th1} to the auxiliary two-disk problem \eqref{eq31}, a series of loop shifting transformations is carried out, in order to convert the plant \eqref{eq29}, \eqref{eq30} into a plant of the form \eqref{eq34} \cite{zhou1996,Safonov1989}. These transformations rely on the following assumptions.
\begin{assumption}\label{ass2}
	\begin{enumerate}
		\item[(i)] The matrix $D_{1, \lambda}$ has full row rank, $\bar{E}_{1} \triangleq D_{1, \lambda} D_{1, \lambda}^{\dagger}>0$.
		\item[(ii)] It holds that
		\begin{equation}
			D_{2, \lambda}(I_{p}-D_{1, \lambda}^{\dagger} \bar{E}_{1}^{-1} D_{1, \lambda}) D_{2, \lambda}^{\dagger}<\left(\gamma^{2}+\lambda^{2}\right) I_{n}. 
		\end{equation}
		\item[(iii)] The following matrix pair is stabilizable:
		\begin{equation}\label{eq52}
			\!\left(A_{\lambda}-B_{\lambda} D_{1, \lambda}^{\dagger} \bar{E}_{1}^{-1} C_{1, \lambda}, B_{\lambda}(I_{p}-D_{1, \lambda}^{\dagger} \bar{E}_{1}^{-1} D_{1, \lambda})\right). 
		\end{equation}
	\end{enumerate}
\end{assumption}

According to Assumption \ref{ass2}(i), the singular value decomposition of $D_{1, \lambda}$ has the form
\begin{equation}\label{eq53}
D_{1, \lambda}=\bar{V}_{12}\left[\begin{array}{ll}
	0 & \bar{\boldsymbol{\Sigma}}_{12} 
\end{array}\right] \bar{W}_{12}^{\dagger},
\end{equation}
where $\bar{W}_{12}, \bar{V}_{12}$ are unitary matrices of dimensions $p \times p$, $n_{y} \times n_{y}$, respectively, and $\bar{\boldsymbol{\Sigma}}_{12}$ is a positive definite diagonal $n_{y} \times n_{y}$ matrix.

Next, let $\bar{\gamma}=\left(\gamma^{2}+\lambda^{2}\right)^{1 / 2}$ and define the matrix
\begin{equation*}
S=\frac{1}{\bar{\gamma}} \bar{W}_{12}^{\dagger}\left(I_{p}-D_{1, \lambda}^{\dagger} \bar{E}_{1}^{-1} D_{1, \lambda}\right) D_{2, \lambda}^{\dagger}.
\end{equation*}
Then it follows from Assumption \ref{ass2}(ii) that
\begin{equation*}
S^{\dagger} S=\frac{1}{\bar{\gamma}^{2}} D_{2, \lambda}\left(I_{p}-D_{1, \lambda}^{\dagger} \bar{E}_{1}^{-1} D_{1, \lambda}\right) D_{2, \lambda}^{\dagger}<I_{n}.
\end{equation*}
Hence, the matrices $R_{1}=I_{n}-S^{\dagger} S, R_{2}=I_{p}-S S^{\dagger}$ are nonsingular, and one can choose Hermitian positive definite $R_{1}^{-1 / 2}, R_{2}^{-1 / 2}$. Also, introduce the singular value decomposition for $R_{1}^{-1 / 2}$,
\begin{equation}
	R_{1}^{-1 / 2}=\bar{W}_{21} \bar{\boldsymbol{\Sigma}}_{21} \bar{W}_{21}^{\dagger},
\end{equation}
where $\bar{W}_{21}$ is a unitary $n \times n$ matrix and $\bar{\boldsymbol{\Sigma}}_{21}>0$ is a diagonal $n \times n$ matrix.

With this notation, define
\begin{align}
	 \tilde{A}=&A_{\lambda}^{\dagger}+\frac{1}{{\bar{\gamma}}^{2}}\left(C_{2, \lambda}^{\dagger}-C_{1, \lambda}^{\dagger} \bar{E}_{1}^{-1} D_{1, \lambda} D_{2, \lambda}^{\dagger}\right) \nonumber \\
	& \times R_{1}^{-1} D_{2, \lambda}\left(I_{p}-D_{1, \lambda}^{\dagger} \bar{E}_{1}^{-1} D_{1, \lambda}\right) B_{\lambda}^{\dagger}, \nonumber \\
	 \tilde{B}_{1}=&\frac{1}{\bar{\gamma}}\left(C_{2, \lambda}^{\dagger}-C_{1, \lambda}^{\dagger} \bar{E}_{1}^{-1} D_{1, \lambda} D_{2, \lambda}^{\dagger}\right) \bar{W}_{21} \bar{\boldsymbol{\Sigma}}_{21},\nonumber \\
	 \tilde{B}_{2}=&C_{1, \lambda}^{\dagger} \bar{V}_{12} \bar{\boldsymbol{\Sigma}}_{12}^{-1},\nonumber \\
	 \tilde{C}_{1}=& R_{2}^{-1 / 2} \bar{W}_{12}^{\dagger} B_{\lambda}^{\dagger},\nonumber \\
	 \tilde{C}_{2}=&\frac{1}{\bar{\gamma}} \bar{W}_{21}^{\dagger} R_{1}^{-1 / 2} D_{2, \lambda}\left(I_{p}-D_{1, \lambda}^{\dagger} \bar{E}_{1}^{-1} D_{1, \lambda}\right) B_{\lambda}^{\dagger},\nonumber \\
	 \tilde{D}_{11}=&0, \;\tilde{D}_{12}=\left[\begin{array}{c}
		0 \\
		I_{n_{y}}
	\end{array}\right], \;\tilde{D}_{21}=I_{n}, \;\tilde{D}_{22}=0 . \label{eq55}
\end{align}

\begin{lemma}\label{lem2}
 Under Assumptions \ref{ass2}(i) and (ii), a controller $\bar{H}_{11}$ which solves the two-disk problem \eqref{eq31} exists if and only if there exists a solution to the two-disk problem \eqref{eq32}, \eqref{eq33} involving the plant
\begin{equation}\label{eq56}
\tilde{\mathscr{P}} \sim\left[\begin{array}{c|cc}
	\tilde{A} & \tilde{B}_{1} & \tilde{B}_{2}  \\
	\hline \tilde{C}_{1} & \tilde{D}_{11} & \tilde{D}_{12} \\
	\tilde{C}_{2} & \tilde{D}_{21} & \tilde{D}_{22}
\end{array}\right]=\left[\begin{array}{c|cc}
	\tilde{A} & \tilde{B}_{1} & \tilde{B}_{2} \\
	\hline \tilde{C}_{1} & 0 & \tilde{D}_{12} \\
	\tilde{C}_{2} & I_{n} & 0
\end{array}\right],
\end{equation}
whose coefficients are given in \eqref{eq55}, the constant $\tilde{\gamma}=1$ and the matrices $V$, $W$ and $J$ defined as
\begin{align}
	V & =\bar{\gamma} \bar{V}_{12} \bar{\boldsymbol{\Sigma}}_{12}^{-1}, \nonumber \\
	W & =\bar{W}_{21} \bar{\boldsymbol{\Sigma}}_{21}^{-1}, \nonumber \\
	J & =\bar{E}_{1}^{-1} D_{1, \lambda} D_{2, \lambda}^{\dagger}. \label{eq57}
\end{align}
If a controller $\mathscr{C}$ solves the latter problem, then the corresponding solution of the problem \eqref{eq31} is given by
\begin{equation}\label{eq58}
	\bar{H}_{11}(s)=\bar{\gamma} \bar{V}_{12} \bar{\boldsymbol{\Sigma}}_{12}^{-1} \mathscr{C}(s) \bar{\boldsymbol{\Sigma}}_{21}^{-1} \bar{W}_{21}^{\dagger}-\bar{E}_{1}^{-1} D_{1, \lambda} D_{2, \lambda}^{\dagger}.
\end{equation}
\end{lemma}
\begin{pf}
The lemma follows as a result of a series of loop shifting transformations applied to the plant $\mathscr{P}$ \eqref{eq29}, \eqref{eq30}; see \cite{zhou1996,Safonov1989}.   \hfill $\Box$
\end{pf}

We now turn to constructing a solution to the coherent equalization under consideration in this paper. For this, we introduce the algebraic Riccati equation
\begin{align}
	& \left(\tilde{A}-\tilde{B}_{2} \tilde{D}_{12}^{\dagger} \tilde{C}_{1}\right)^{\dagger} \tilde{X}+\tilde{X}\left(\tilde{A}-\tilde{B}_{2} \tilde{D}_{12}^{\dagger} \tilde{C}_{1}\right) \nonumber\\
	& +\tilde{X}\left(\tilde{B}_{1} \tilde{B}_{1}^{\dagger}-\tilde{B}_{2} \tilde{B}_{2}^{\dagger}\right) \tilde{X}+\tilde{C}_{1}^{\dagger}\left(I_{p}-\tilde{D}_{12} \tilde{D}_{12}^{\dagger}\right) \tilde{C}_{1}=0. \label{eq59}
\end{align}
Associated with the stabilizing nonnegative definite solution $\tilde{X}$ to this equation, introduce coefficient matrices for the central controller $K_{c}$ in \eqref{eq36}:
\begin{align}
	A_{K_{c}} & =\tilde{A}-\tilde{B}_{1} \tilde{C}_{2}-\tilde{B}_{2}\left(\tilde{B}_{2}^{\dagger} \tilde{X}+\tilde{D}_{12}^{\dagger} \tilde{C}_{1}\right) \nonumber \\
	& =A_{\lambda}^{\dagger}-\tilde{B}_{2}\left(\tilde{B}_{2}^{\dagger} \tilde{X}+\tilde{D}_{12}^{\dagger} \tilde{C}_{1}\right),  \nonumber \\
	B_{K_{c}, 1} & =\tilde{B}_{1},  \nonumber \\
	B_{K_{c}, 2} & =\tilde{B}_{2},  \nonumber \\
	C_{K_{c}, 1} & =-\left(\tilde{B}_{2}^{\dagger} \tilde{X}+\tilde{D}_{12}^{\dagger} \tilde{C}_{1}\right),  \nonumber \\
	C_{K_{c}, 2} & =-\left(\tilde{C}_{2}+\tilde{B}_{1}^{\dagger} \tilde{X}\right). \label{eq60}
\end{align}
The following theorem is the main theoretical result of the paper. It underpins the algorithm for the synthesis of coherent passive equalizers which will be presented in the next subsection.

\begin{theorem}\label{th2}
Given a $\gamma>0$, suppose Assumptions \ref{ass1} and \ref{ass2} are satisfied and the matrix pair
\begin{equation}\label{eq61}
	\Big(A_{\lambda}+\frac{1}{\bar{\gamma}^{2}} B_{\lambda}(I_{p}-{D}_{1, \lambda}^{\dagger} \bar{E}_{1}^{-1} {D}_{1, \lambda}) D_{2, \lambda}^{\dagger} R_{1}^{-1} C_{2, \lambda}, C_{1, \lambda}\Big) 
\end{equation}
with $\bar{\gamma}=\left(\gamma^{2}+\lambda^{2}\right)^{1 / 2}$, is detectable. Also, suppose the Riccati equation \eqref{eq59} has the stabilizing solution $\tilde{X} \geq 0$. Furthermore, suppose that the central controller $K_{c}$ defined in \eqref{eq60} satisfies the conditions of Theorem \ref{th1} with $\tilde{\gamma}=1$. Obtain $\bar{H}_{11}$ using equation \eqref{eq58}, in which $\mathscr{C}$ is defined in \eqref{eq47}. Then the transfer function $H(s)$, composed of
\begin{equation}\label{eq62}
	H_{11}(s)=\bar{H}_{11}\left(s^{*}\right)^{\dagger},
\end{equation}
and $H_{12}(s), H_{21}(s), H_{22}(s)$ defined in \eqref{eq15}, \eqref{eq16} represents a completely passive physically realizable equalizer which guarantees that
\begin{equation}\label{eq63}
	\bar{\boldsymbol{\sigma}}\left(P_{e}(i \omega)\right)<\gamma^{2} \quad \forall \omega \in \bar{\mathbf{R}}.
\end{equation}
\end{theorem}
\begin{pf}
First we note that if a rational transfer function $\bar{H}_{11} \in H_{\infty}$ solves the two-disk problem \eqref{eq31} then $H_{11}(s)=\bar{H}_{11}\left(s^{*}\right)^{\dagger} \in \mathscr{H}_{11, \gamma}^{-}$. This follows from Lemma \ref{lem1} and the contractiveness of $\bar{H}_{11}$; see the second condition in \eqref{eq31}. Theorem 1 in \cite{Ugrinovskii2024a} establishes that the transfer function $H(s)$ constructed from such $H_{11} \in \mathscr{H}_{11, \gamma}^{-}$ and the blocks $H_{12}$, $H_{21}$ and $H_{22}$ defined in \eqref{eq15}, \eqref{eq16} meets the requirements of Definition \ref{def1}. Also, according to Lemma \ref{lem1}, the first norm condition in \eqref{eq31} is equivalent to \eqref{eq12}, and the latter is trivially equivalent to \eqref{eq63}.

Thus, to prove the statement of the theorem it suffices to show that the transfer function $\bar{H}_{11} \in {H}_{\infty}$ defined by \eqref{eq58} via $\mathscr{C}$ in \eqref{eq47} solves the two-disk problem \eqref{eq31}. Due to Lemma \ref{lem2}, proving this amounts to proving that the controller $\mathscr{C}$ defined in \eqref{eq47} solves the two-disk problem \eqref{eq32}, \eqref{eq33} in which $\tilde{\gamma}=1$. The latter claim follows from Theorem \ref{th1}, provided the plant \eqref{eq56}, \eqref{eq55} satisfies conditions (A1)-(A4) in the previous section. Indeed, under these conditions, the existence of the stabilizing solution $\tilde{X} \geq 0$ to the Riccati equation \eqref{eq59} is equivalent to the satisfaction of the $H_{\infty}$ norm condition \eqref{eq32} by any controller $\mathscr{C}$ which has the LFT structure \eqref{eq35}, with $K_{c}$ being the central controller \eqref{eq36}, \eqref{eq60} and $\tilde{\gamma}=1$ \cite{zhou1996}. Normally, this also requires the existence of the stabilizing solution to the Riccati equation
\begin{align}
	& \left(\tilde{A}-\tilde{B}_{1} \tilde{C}_{2}\right) \tilde{Y}+\tilde{Y}\left(\tilde{A}-\tilde{B}_{1} \tilde{C}_{2}\right)^{\dagger} \nonumber\\
	& \quad+\tilde{Y}\left(\tilde{C}_{1}^{\dagger} \tilde{C}_{1}-\tilde{C}_{2}^{\dagger} \tilde{C}_{2}\right) \tilde{Y}+\tilde{B}_{1}\left(I_{n}-\tilde{D}_{21}^{\dagger} \tilde{D}_{21}\right) \tilde{B}_{1}^{\dagger}=0. \label{eq64}
\end{align}
Also, the spectral radius of $\tilde{Y} \tilde{X}$ must be less than $1$. However, these requirements are met trivially in our case with $\tilde{Y}=0$. This fact is a by-product of the disturbance feedforward structure of the plant $\mathscr{P}$. Indeed, the substitution of \eqref{eq55} into \eqref{eq64} reduces this Riccati equation to
\begin{equation}
	A_{\lambda}^{\dagger} \tilde{Y}+\tilde{Y} A_{\lambda}+\tilde{Y} B_{\lambda} B_{\lambda}^{\dagger} \tilde{Y}=0,
\end{equation}
since $I_{n}-\tilde{D}_{21}^{\dagger} \tilde{D}_{21}=0$ and
\begin{equation}\label{eq66}
	\tilde{A}-\tilde{B}_{1} \tilde{C}_{2}=A_{\lambda}^{\dagger}, \quad \tilde{C}_{1}^{\dagger} \tilde{C}_{1}-\tilde{C}_{2}^{\dagger} \tilde{C}_{2}=B_{\lambda} B_{\lambda}^{\dagger}.
\end{equation}
Therefore $\tilde{Y}=0$ is the stabilizing solution to \eqref{eq64} since $A_{\lambda}+B_{\lambda} B_{\lambda}^{\dagger} \tilde{Y}=A_{\lambda}$ is a Hurwitz matrix, according to Assumption \ref{ass1}. Also, the spectral radius of $\tilde{Y} \tilde{X}$ is zero in this case. The claim of the theorem now follows from Theorem \ref{th1} since according to that theorem, the parameter $\mathscr{Q}=D_{Q}$ constructed from the convex LMI problem \eqref{eq38}-\eqref{eq41} yields $\mathscr{C}$ in \eqref{eq47} which satisfies \eqref{eq33}.

Thus, to complete the proof, it remains to establish that the plant \eqref{eq56}, \eqref{eq55} satisfies conditions (A1)-(A4) of Theorem \ref{th1}. The last line in \eqref{eq55} shows that (A2) is indeed satisfied. To validate the remaining conditions (A1), (A3) and (A4), we use a series of propositions. Note that the detectability of $(\tilde{C}_{2}, \tilde{A})$ follows immediately from \eqref{eq66} since $A_{\lambda}$ is Hurwitz.
\begin{proposition}\label{pro1}
Under Assumption \ref{ass1}, (A4) is satisfied.
\end{proposition}
\begin{pf}
 Since by Assumption \ref{ass1}, $A_{\lambda}^{\dagger}$ is a stable matrix, the pair $(A_{\lambda}^{\dagger}, \tilde{B}_{1})$ is stabilizable, i.e., the matrix $\left[A_{\lambda}^{\dagger}-s I_{m}\;\; \tilde{B}_{1}\right]$ has full row rank for all $s$, $\operatorname{Re} s \geq 0$. Consequently, $\left[\begin{array}{cc}A_{\lambda}^{\dagger}-s I_{m} & \tilde{B}_{1} \\ 0 & I_{n}\end{array}\right]$ has full row rank for all $s$, $\operatorname{Re} s \geq 0$. Therefore, using the first identity \eqref{eq66}, we conclude that for any such $s$
\begin{equation*}
\!\left[\begin{array}{cc}\tilde{A}-s I_{m} & \tilde{B}_{1} \\ \tilde{C}_{2} & I_{n}\end{array}\right]\!\!\!\left[\begin{array}{l}x \\ \varpi\end{array}\right]\!\!=\!\!\left[\begin{array}{cc}A_{\lambda}^{\dagger}-s I_{m} & \tilde{B}_{1} \\ 0 & I_{n}\end{array}\right]\!\!\!\left[\begin{array}{ll}I_{m} & 0 \\ \tilde{C}_{2} & I_{n}\end{array}\right]\!\!\left[\begin{array}{l}x \\ \varpi\end{array}\right]\!\!=\!0
\end{equation*}
if and only if $\left[\begin{array}{cc}I_{m} & 0 \\ \tilde{C}_{2} & I_{n}\end{array}\right]\left[\begin{array}{c}x \\ \varpi\end{array}\right]=0$. The latter equation implies that $x=0, \varpi=0$. Thus, (A4) holds true.   \hfill $\Box$
\end{pf}

The next proposition validates (A1). Note that the detectability of $(\tilde{C}_{2}, \tilde{A})$ follows immediately from \eqref{eq66} since $A_{\lambda}$ is Hurwitz. Therefore, we only need to validate the stabilizability of $(\tilde{A}, \tilde{B}_{2})$.
\begin{proposition}\label{pro2}
	If the matrix pair \eqref{eq61} is detectable, then $(\tilde{A}, \tilde{B}_{2})$ is stabilizable.
\end{proposition}
\begin{pf}
 Let $K$ be a matrix such that
 \begin{equation*}
 	\begin{aligned}
A^{K} \triangleq& A_{\lambda}+\frac{1}{\bar{\gamma}^{2}} B_{\lambda}\left(I_{p}-{D}_{1, \lambda}^{\dagger} \bar{E}_{1}^{-1} {D}_{1, \lambda}\right) D_{2, \lambda}^{\dagger} R_{1}^{-1} C_{2, \lambda}\\&+K C_{1, \lambda}
	\end{aligned}
 \end{equation*}
is a stable matrix. Also, let
\begin{equation*}
\begin{aligned}
	\tilde{K}= & \Big(K+\frac{1}{\bar{\gamma}^{2}} B_{\lambda}(I_{p}-{D}_{1, \lambda}^{\dagger} \bar{E}_{1}^{-1} {D}_{1, \lambda}) \\
	& \times D_{2, \lambda}^{\dagger} R_{1}^{-1} D_{2, \lambda} D_{1, \lambda}^{\dagger} \bar{E}_{1}^{-1}\Big) \bar{V}_{12} \bar{\boldsymbol{\Sigma}}_{12}.
\end{aligned}
\end{equation*}
Then it holds that $\tilde{A}^{\dagger}+\tilde{K} \tilde{B}_{2}^{\dagger}=A^{K}$ is stable. Thus, $(\tilde{A}, \tilde{B}_{2})$ is stabilizable.   \hfill $\Box$
\end{pf}

\begin{proposition}\label{pro3}
 Under Assumption \ref{ass2}(iii), (A3) holds true.
\end{proposition}
\begin{pf}
 Assumption \ref{ass2}(iii) is equivalent to the detectability of the pair
\begin{equation}\label{eq67}
	\left(A_{\lambda}^{\dagger}-C_{1, \lambda}^{\dagger} \bar{E}_{1}^{-1} D_{1, \lambda} B_{\lambda}^{\dagger},\left(I_{p}-D_{1, \lambda}^{\dagger} \bar{E}_{1}^{-1} D_{1, \lambda}\right) B_{\lambda}^{\dagger}\right).
\end{equation}
The latter property holds if and only if the matrix $\left[\begin{array}{cc}A_{\lambda}^{\dagger}-s I_{m} & C_{1, \lambda}^{\dagger} \\ B_{\lambda}^{\dagger} & D_{1, \lambda}^{\dagger}\end{array}\right]$ has full column rank for all $s, \operatorname{Re} s \geq 0$ \cite[Lemma 13.9]{zhou1996}\footnote{In \cite[Lemma 13.9]{zhou1996}, the proof is given for the case where the matrix pair \eqref{eq67} has no unobservable modes on $i \omega$-axis. In the case where the matrix pair \eqref{eq67} is detectable, the proof follows along the same lines, except that $i \omega$ is to be replaced by $s$ with $\operatorname{Re} s \geq 0$.}. We now use this property to prove that
\begin{equation}\label{eq68}
\left[\begin{array}{cc}
	A_{\lambda}^{\dagger}-s I_{m} & \tilde{B}_{2}  \\
	\bar{W}_{12}^{\dagger} B_{\lambda}^{\dagger} & \tilde{D}_{12}
\end{array}\right]
\end{equation}
also has full column rank for all $s, \operatorname{Re} s \geq 0$. This is readily seen from the equation
\begin{align}
	& {\left[\begin{array}{cc}
			A_{\lambda}^{\dagger}-s I_{m} & \tilde{B}_{2} \\
			\bar{W}_{12}^{\dagger} B_{\lambda}^{\dagger} & \tilde{D}_{12}
		\end{array}\right]\left[\begin{array}{l}
			x \\
			v
		\end{array}\right]} \nonumber\\
	 =&\left[\begin{array}{cc}
		I_{m} & 0 \\
		0 & \bar{W}_{12}^{\dagger}
	\end{array}\right]\left[\begin{array}{cc}
		A_{\lambda}^{\dagger}-s I_{m} & C_{1, \lambda}^{\dagger} \\
		B_{\lambda}^{\dagger} & D_{1, \lambda}^{\dagger}
	\end{array}\right]\left[\begin{array}{cc}
		I_{m} & 0 \\
		0 & \bar{V}_{12} {\bar{\boldsymbol{\Sigma}}}_{12}^{-1}
	\end{array}\right]\left[\begin{array}{l}
		x \\
		v
	\end{array}\right] \nonumber\\
	 =&0. \label{eq69}
\end{align}
In the second line, we used \eqref{eq53} according to which $\tilde{D}_{12}=$ $\bar{W}_{12}^{\dagger} D_{1, \lambda}^{\dagger} V_{12} \bar{\boldsymbol{\Sigma}}_{12}^{-1}$. Since all three matrix factors in the second line are nonsingular when $\operatorname{Re} s \geq 0$, then $x=0$, $v=0$ is the only solution to this equation. That is, the matrix \eqref{eq68} has full column rank for all such $s$.

Next, fix $s$ such that $\operatorname{Re} s \geq 0$ and consider the equations
\begin{align}
	& \left(\tilde{A}-s I_{m}\right) x+\tilde{B}_{2} v=0, \label{eq70}\\
	& \tilde{C}_{1} x+\tilde{D}_{12} v=0. \label{eq71}
\end{align}
These equations can be written as
\begin{equation}\label{eq72}
	\begin{aligned}
(A_{\lambda}^{\dagger}-s I_{m}) x&+\tilde{B}_{2} v+\frac{1}{\bar\gamma^{2}}(C_{2, \lambda}^{\dagger}-C_{1, \lambda}^{\dagger} \bar{E}_{1}^{-1} D_{1, \lambda} D_{2, \lambda}^{\dagger})\\
&\times S^{\dagger} R_{2}^{-1} \bar{W}_{12}^{\dagger} B_{\lambda}^{\dagger} x=0,\\
	\end{aligned}
\end{equation}
\begin{equation}\label{eq73}
R_{2}^{-1 / 2} \bar{W}_{12}^{\dagger} B_{\lambda}^{\dagger} x+\tilde{D}_{12} v=0.
\end{equation}

Note the following identities
\begin{equation*}
	\begin{aligned}
		& I_{p}-D_{1, \lambda}^{\dagger} \bar{E}_{1}^{-1} D_{1, \lambda}=\bar{W}_{12}\left[\begin{array}{cc}
			I_{p-n_{y}} & 0 \\
			0 & 0
		\end{array}\right] \bar{W}_{12}^{\dagger}, \\
		& S=\frac{1}{\bar{\gamma}}\left[\begin{array}{cc}
			I_{p-n_{y}} & 0 \\
			0 & 0
		\end{array}\right] \bar{W}_{12}^{\dagger} D_{2, \lambda}^{\dagger} \triangleq\left[\begin{array}{c}
			\Delta \\
			0
		\end{array}\right], \\
		& R_{2}=\left[\begin{array}{cc}
			I_{p-n_{y}}-\Delta \Delta^{\dagger} & 0 \\
			0 & I_{n_{y}}
		\end{array}\right], \\
		& S^{\dagger} R_{2}^{-1 / 2} \tilde{D}_{12}=\left[\Delta^{\dagger}\left(I_{p-n_{y}}-\Delta \Delta^{\dagger}\right)^{-1 / 2}\;\; 0\right]\left[\begin{array}{c}
			0 \\
			I_{n_{y}}
		\end{array}\right]=0.
	\end{aligned}
\end{equation*}
Then, after multiplying \eqref{eq73} by $S^{\dagger} R_{2}^{-1 / 2}$ from the left, we obtain $S^{\dagger} R_{2}^{-1} \bar{W}_{12}^{\dagger} B_{\lambda}^{\dagger} x=0$. After substituting this into \eqref{eq72} and also multiplying \eqref{eq73} by $R_{2}^{1 / 2}$ from the left, we obtain
\begin{equation*}
	\begin{aligned}
		& (A_{\lambda}^{\dagger}-s I_{m}) x+\tilde{B}_{2} v=0, \\
		& \bar{W}_{12}^{\dagger} B_{\lambda}^{\dagger} x+R_{2}^{1 / 2} \tilde{D}_{12} v=\bar{W}_{12}^{\dagger} B_{\lambda}^{\dagger} x+\tilde{D}_{12} v=0.
	\end{aligned}
\end{equation*}
In the last equation, we have used the identity $R_{2}^{1 / 2} \tilde{D}_{12}=$ $\tilde{D}_{12}$. Thus we conclude that any solution to \eqref{eq70}, \eqref{eq71} is also a solution to equation \eqref{eq69}. However since we have shown that the matrix \eqref{eq68} has full column rank for all $s$, $\operatorname{Re} s \geq 0$, this implies that $x=0, v=0$ is the unique solution to \eqref{eq70}, \eqref{eq71}. This shows that (A3) holds true.    \hfill $\Box$
\end{pf}

Propositions \ref{pro1}-\ref{pro3} validate conditions (A1)-(A4) needed to apply Theorem \ref{th1}. This concludes the proof of Theorem \ref{th2}. \hfill $\Box$
\end{pf}

We remark that \eqref{eq63} is an upper bound on the optimal mean-square equalization error \eqref{eq2}. This upper bound can be optimized, e.g., using the bisection method since feasible values of $\gamma^{2}$ lie in the interval $\left[0, \bar{\boldsymbol{\sigma}}\left(\Sigma_{u}^{T}+2 I_{n}\right)\right]$. The upper boundary of this interval corresponds to the trivial equalizer $H(s) = \begin{bmatrix}0 &I\\I &0\end{bmatrix}$; see \cite{Ugrinovskii2024}. The gap between this
optimized $\gamma^{2}$ and the optimal value \eqref{eq2} can be estimated using the following semidefinite program in variables $\nu^{2}$ and $\left\{H_{11, l}, l=1, \ldots, L\right\}$:
\begin{align}
	\nu_{\bar{\omega}}^{2} & =\inf \nu^{2}   \label{eq74}\\
	\text { s.t. } & {\left[\begin{array}{cc}
			\Xi_{11, l} & H_{11, l} M\left(i \omega_{l}\right) \\
			M\left(i \omega_{l}\right)^{\dagger} H_{11, l}^{\dagger} & -I_{q}
		\end{array}\right] \leq 0, } \nonumber \\
	& {\left[\begin{array}{cc}
			I_{n} & H_{11, l} \\
			H_{11, l}^{\dagger} & I_{n}
		\end{array}\right] \geq 0 \quad \forall l=1, \ldots, L. } \nonumber
\end{align}
Here, $\bar{\omega} \triangleq\left\{\omega_{l}, l=1, \ldots, L\right\}$ is an arbitrary finite grid of frequencies,
\begin{equation*}
	\begin{aligned}
		\Xi_{11, l} \triangleq & \left(2-\nu^{2}\right) I+\Sigma_{u}^{T}-H_{11, l} G_{11}\left(i \omega_{l}\right)\left(I+\Sigma_{u}^{T}\right) \\
		& -\left(I+\Sigma_{u}^{T}\right) G_{11}\left(i \omega_{l}\right)^{\dagger} H_{11, l}^{\dagger},
	\end{aligned}
\end{equation*}
$M(s)$ is a stable $n_{y} \times q$ spectral factor of $\Psi(s)$ which is analytic in the complex half-plane $\operatorname{Re} s>-\tau\;(\exists \tau>0)$:
\begin{equation*}
	\Psi(s)=M(s) M(s)^{H}.
\end{equation*}
\begin{theorem}\label{th3}
For any $\gamma>0$ such that $\mathscr{H}_{11, \gamma}^{-} \neq \emptyset$,
\begin{equation}\label{eq75}
	\sup _{\bar{\omega}} \nu_{\bar{\omega}}^{2} \leq \inf \sup _{\omega} \bar{\boldsymbol{\sigma}}\left(P_{e}(i \omega)\right)<\gamma^{2}. 
\end{equation}	
\end{theorem}
\begin{pf}
 Let $H_{11}$ be an arbitrary element of $\mathscr{H}_{11, \gamma}^{-}$, then the right-hand side inequality in \eqref{eq75} follows from \eqref{eq12}, since each $H_{11} \in \mathscr{H}_{11, \gamma}^{-}$ generates a physically realizable $H(s)$, as discussed previously.

To prove the inequality on the left-hand side of \eqref{eq75}, select a physically realizable $H(s)$ and a finite grid of frequencies $\bar{\omega}=\left\{\omega_{l}, l=1, \ldots, L\right\}$. Consider the (1,1) block of $H(s)$ and the corresponding PSD matrix $P_{e}(i \omega)$. It then holds that
\begin{equation}\label{eq76}
	\tilde{\nu}_{\bar{\omega}} \triangleq \max _{\omega_{l} \in \bar{\omega}} \bar{\boldsymbol{\sigma}}\left(P_{e}\left(i \omega_{l}\right)\right) \leq \sup _{\omega} \bar{\boldsymbol{\sigma}}\left(P_{e}(i \omega)\right) . 
\end{equation}
Using the Schur complement, the definition of $\tilde{\nu}_{\bar{\omega}}$ implies that
\begin{equation*}
\begin{aligned}
	& {\left[\begin{array}{cc}
			\tilde{\Xi}_{11}\left(i \omega_{l}\right) & H_{11}\left(i \omega_{l}\right) M\left(i \omega_{l}\right) \\
			M\left(i \omega_{l}\right)^{\dagger} H_{11}\left(i \omega_{l}\right)^{\dagger} & -I_{q}
		\end{array}\right] \leq 0,} \\
	&\quad \forall l=1, \ldots, L,
\end{aligned}
\end{equation*}
where
\begin{equation*}
	\begin{aligned}
		\tilde{\Xi}_{11}(s) \triangleq & \left(2-\tilde{\nu}_{\bar{\omega}}^{2}\right) I+\Sigma_{u}^{T}-H_{11}(s) G_{11}(s)\left(I+\Sigma_{u}^{T}\right) \\
		& -\left(I+\Sigma_{u}^{T}\right) G_{11}(s)^{H} H_{11}(s)^{H}.
	\end{aligned}
\end{equation*}
Also, it follows from \eqref{eq6} that
\begin{equation*}
\left[\begin{array}{cc}
	I_{n} & H_{11}\left(i \omega_{l}\right) \\
	H_{11}\left(i \omega_{l}\right)^{\dagger} & I_{n}
\end{array}\right] \geq 0 \quad \forall l=1, \ldots, L.
\end{equation*}
The last two inequalities show that the collection $\left\{\tilde{\nu}_{\bar{\omega}}, H_{11}\left(i \omega_{1}\right), \ldots, H_{11}\left(i \omega_{L}\right)\right\}$ belongs to the feasible set of the SDP \eqref{eq74}. Therefore, combining this observation with \eqref{eq76} we conclude
\begin{equation}\label{eq77}
	\nu_{\bar{\omega}} \leq \tilde{\nu}_{\bar{\omega}} \leq \sup _{\omega} \bar{\boldsymbol{\sigma}}\left(P_{e}(i \omega)\right).
\end{equation}
The quantity on the left holds for any selected $H(s)$ and the quantity on the right holds irrespective of the selected grid $\bar{\omega}$. Therefore, the inequality will be preserved after we infimize the left-hand side of \eqref{eq77} with respect to $\bar{\omega}$, and after that, infimize the right-hand side of \eqref{eq77} with respect to physically realizable $H(s)$. This will yield the inequality on the left-hand side of \eqref{eq75}.   \hfill $\Box$
\end{pf}

Theorem \ref{th3} provides a tractable quantitative measure of the gap between \eqref{eq2} and the optimized $\gamma^{2}$, since both the optimized $\gamma^{2}$ and $\nu_{\bar{\omega}}^2$ are computable, one from Theorem \ref{th2}, and another from \eqref{eq74}.

\subsection{An algorithm for synthesis of a coherent equalizer}
Theorem \ref{th2} is constructive in the sense that an algorithm for the synthesis of a coherent equalizer can be derived from it. The first part of the algorithm is to construct an $H_{11} \in$ $\mathscr{H}_{11, \gamma}^{-}$. This $H_{11}$ is then used to compute the remaining blocks of $H(s)$. This two-part structure is analogous to the structure of the algorithm proposed in \cite{Ugrinovskii2024a}. In fact, both papers use the same procedure for constructing $H_{12}, H_{21}$ and $H_{22}$; see Steps 4 and 5 of Algorithm 1 presented below. However, Steps 1-3 are different in that in the algorithm proposed below $H_{11} \in \mathscr{H}_{11, \gamma}^{-}$ is derived from the two-disk problem \eqref{eq31}. In contrast, the $H_{\infty}$ problem in \cite{Ugrinovskii2024a} involved only the first norm condition in \eqref{eq31}, and \cite{Ugrinovskii2024a}  relied on condition \eqref{eq19} to ascertain that the second condition in \eqref{eq31} was also satisfied.

\section*{Algorithm 1}
\begin{enumerate}
	\item Select $\lambda \geq 0$ which satisfies Assumption \ref{ass1} and compute the minimal realization of the spectral factor $\Upsilon_{\lambda}(s)$ of $\Phi_{\lambda}(s)$. Also, select a $\gamma>0$, compute $\bar{\gamma}=\left(\gamma^{2}+\lambda^{2}\right)^{1 / 2}$ and validate Assumption \ref{ass2}. Then construct the matrices $\tilde{A}, \tilde{B}_{1}, \tilde{B}_{2}, \tilde{C}_{1}$, and $\tilde{C}_{2}$ according to \eqref{eq55}. Obtain the stabilizing solution $\tilde{X}$ to the Riccati equation \eqref{eq59}.
	
	\item Using \eqref{eq60} construct the coefficients of the central controller $K_{c}$ in \eqref{eq36} and compute feasible solutions $X_{1}$ and $Y_{1}$ to the LMIs \eqref{eq38}-\eqref{eq41}. Using the found matrices $X_{1}, Y_{1}$, construct an $m \times m$ matrix $X_{2}$ such that $X_{2} X_{2}^{\dagger}=X_{1}-Y_{1}^{-1}$. Such $X_{2}$ exists since $X_{1}-Y_{1}^{-1} \geq 0$ according to \eqref{eq41}. Then construct the $2 m \times 2 m$ matrix $\hat{X}$, defined in \eqref{eq42}. Note that $\hat{X}>0$, since by definition of $X_{2}, X_{1}-X_{2} X_{2}^{\dagger}=Y_{1}^{-1}>0$; see \eqref{eq38}.
	
	\item With this $\hat{X}$ and $\tilde{\gamma}=1$, solve the linear matrix inequalities \eqref{eq44}, \eqref{eq45} for $Q$ defined in \eqref{eq46} and construct $\mathscr{C}$ via \eqref{eq47}. Then construct the transfer function $\bar{H}_{11}(s)$ in \eqref{eq58} and obtain $H_{11}(s)=\bar{H}_{11}\left(s^{*}\right)^{\dagger}$ as per \eqref{eq62}. It is shown in the proof of Theorem \ref{th2} that the transfer function $H_{11}(s)$ constructed this way solves the auxiliary problem described in Definition \ref{def2}; i.e., $H_{11} \in \mathscr{H}_{11, \gamma}^{-}$.
	
	\item This step and the next step follow the corresponding steps of the algorithm proposed in \cite{Ugrinovskii2024a}. Using the found $H_{11}(s)$ obtain the transfer functions $Z_{1}(s), Z_{2}(s)$ in \eqref{eq14}, then compute their spectral factors $H_{12}(s)$ and $\tilde{H}_{21}(s)$ as per equations \eqref{eq15} and \eqref{eq17}; see \cite{Ugrinovskii2024a} for a suitable choice of such spectral factors.
	
	\item Obtain the remaining transfer functions $H_{21}(s)$ and $H_{22}(s)$ using equations \eqref{eq16}.
	
	\item Optionally, select a frequency grid $\bar{\omega}=\left\{\omega_{l}, l=\right.$ $1, \ldots, L\}$ and solve the semidefinite program \eqref{eq74}, to estimate the gap between $\gamma^{2}$ and \eqref{eq2}. If necessary, reduce $\gamma^{2}$ and/or select a different grid $\bar{\omega}$, then repeat the above steps until $\gamma^{2}-\nu_{\bar{\omega}}^{2}$ is minimized.
	
\end{enumerate}

\section{Illustrative example}\label{sec5}
\begin{figure}
	\centering
	\includegraphics[width=1\columnwidth]{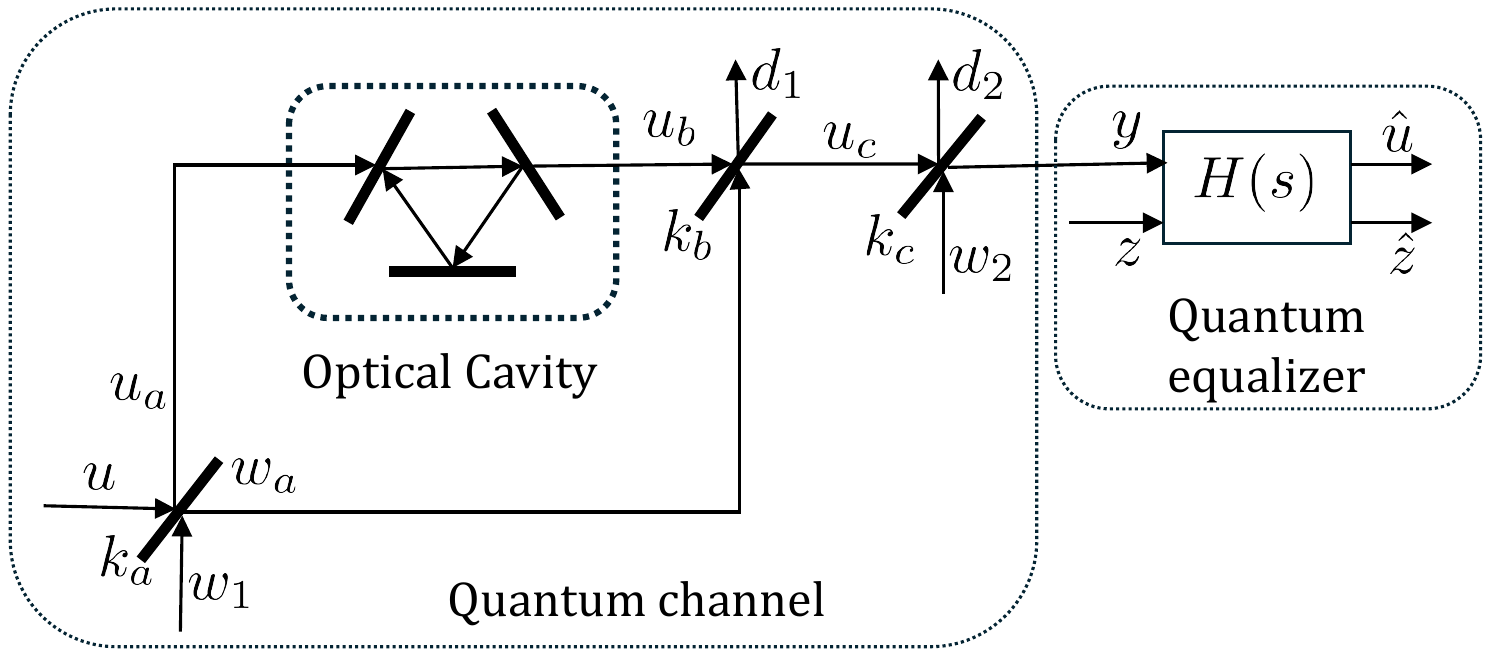}
	\caption{A cavity, beam splitters and an equalizer system from \cite{Ugrinovskii2024a}.\label{fig3}}
\end{figure}

To illustrate the synthesis procedure described in the previous sections and compare it with the previous methods, we consider the quantum optical equalization system shown in Fig. \ref{fig3}. This system was also used as an example in \cite{Ugrinovskii2024a}. 
The channel consists of an optical cavity and three optical beam splitters, forming a passive quantum system that can be implemented in experiments~\cite{hendra2017}. It is similar to the optical cavity system analyzed in~\cite{Ugrinovskii2024} but includes an additional beam splitter. The third beam splitter introduces environmental noise analogous to additive Gaussian white noise in conventional communication channels~\cite{hassibi1999indefinite}. This additional element allows us to demonstrate the application of both our synthesis method and the approach in~\cite{Ugrinovskii2024a} in a low signal-to-noise ratio scenario. As in classical systems, the low signal-to-noise condition arises when \( \sigma_{w_2}^2 \) is sufficiently large.

The input operator $u$ in this example is scalar, and $\Sigma_{u}$ is a real constant. To emphasize this, we use the notation $\Sigma_{u}=\sigma_{u}^{2}$. The environment is represented by the quantum noises $w_{1}$, $w_{2}$, thus $w=\operatorname{col}\left(w_{1}, w_{2}\right)$ consists of two scalar operators. We assume that $\Sigma_{w}=\left[\begin{array}{cc}\sigma_{w_{1}}^{2} & 0 \\ 0 & \sigma_{w_{2}}^{2}\end{array}\right]$. As in \cite{Ugrinovskii2024a}, we assume that the transmittance parameters $k_{a}^{2}, k_{b}^{2}, k_{c}^{2}$ of the beam splitters are real positive numbers and that $k_{a}, k_{b}$ and $k_{c}$ are also real positive constants, and that $k_{a}=k_{b}=k$. Thus, the relations between the inputs and outputs of the beam splitters are
\begin{equation*}
	\begin{aligned}
		& {\left[\begin{array}{c}
				u_{a} \\
				w_{a}
			\end{array}\right]=\left[\begin{array}{cc}
				k & l \\
				-l & k
			\end{array}\right]\left[\begin{array}{c}
				u \\
				w_{1}
			\end{array}\right], \quad\left[\begin{array}{c}
				u_{c} \\
				d_{1}
			\end{array}\right]=\left[\begin{array}{cc}
				k & l \\
				-l & k
			\end{array}\right]\left[\begin{array}{c}
				u_{b} \\
				w_{a}
			\end{array}\right]} \\
		& {\left[\begin{array}{c}
				y \\
				d_{2}
			\end{array}\right]=\left[\begin{array}{cc}
				k_{c} & l_{c} \\
				-l_{c} & k_{c}
			\end{array}\right]\left[\begin{array}{c}
				u_{c} \\
				w_{2}
			\end{array}\right]},
	\end{aligned}
\end{equation*}
where $l \triangleq \sqrt{1-k^{2}}, l_{c} \triangleq \sqrt{1-k_{c}^{2}}$ are real positive numbers. The transfer function of the optical cavity is $G_{c}(s)=\frac{s-\kappa+i \Omega}{s+\kappa+i \Omega}$, i.e., $u_{b}=G_{c}(s) u_{a} ; \kappa>0$, $\Omega$ are real numbers. Then the elements of the transfer function $G(s)$ of the channel are
\begin{align}
	& G_{11}(s)=k_{c}\left(k^{2} G_{c}(s)-\left(1-k^{2}\right)\right), \nonumber\\
	& G_{12}(s)=\left[\begin{array}{ll}
		k_{c} k \sqrt{1-k^{2}}\left(G_{c}(s)+1\right) & \sqrt{1-k_{c}^{2}}
	\end{array}\right], \nonumber\\
	& G_{21}(s)=\left[\begin{array}{cc}
		-k \sqrt{1-k^{2}}\left(G_{c}(s)+1\right)\\
		-\sqrt{1-k_{c}^{2}}\left(k^{2} G_{c}(s)-\left(1-k^{2}\right)\right)
	\end{array}\right], \nonumber \\
	& G_{22}(s)=\left[\begin{array}{cc}
		k^{2}-\left(1-k^{2}\right) G_{c}(s) & 0 \\
		-\sqrt{1-k_{c}^{2}} k \sqrt{1-k^{2}}\left(G_{c}(s)+1\right) & k_{c}
	\end{array}\right].
\end{align}
We adopt the same assumptions as in \cite{Ugrinovskii2024a}: $\sigma_{w_{1}}^{2}>\sigma_{u}^{2}>0$ and $k^{2}<\frac{1}{2}$. Under these assumptions,
\begin{equation*}
\begin{aligned}
	& \rho \triangleq 1+\frac{\sigma_{u}^{2}}{2\left(\sigma_{w_{1}}^{2}-\sigma_{u}^{2}\right) k^{2}\left(1-k^{2}\right)}>1, \\
	& \hat{\rho} \triangleq \frac{\rho-1}{\rho+1} \in(0,1), \\
	& \delta \triangleq \frac{\sqrt{1-k^{2}}}{k}>1, \quad \hat{\delta} \triangleq \frac{\delta^{2}+1}{\delta^{2}-1}=\frac{1}{1-2 k^{2}}>1.
\end{aligned}
\end{equation*}
Using these notations, the function $\Psi(s)$ given in equation \eqref{eq10} is expressed as
\begin{equation}
\Psi(s)=k_{c}^{2} \mu^{2} \frac{(s+i \Omega)^{2}-\hat{\rho} \kappa^{2}}{(s+i \Omega)^{2}-\kappa^{2}}+\left(1-k_{c}^{2}\right) \sigma_{w_{2}}^{2},
\end{equation}
where $\mu \triangleq \sqrt{2\left(\sigma_{w_{1}}^{2}-\sigma_{u}^{2}\right) k^{2}\left(1-k^{2}\right)(1+\rho)}$. This gives the expression for the matrix $\Phi(s)$ in equation \eqref{eq9},
\begin{equation*}
\begin{aligned}
	& \Phi(s) \\
	 =&\left[\begin{array}{cc}
		k_{c}^{2} \mu^{2} \frac{(s+i \Omega)^{2}-\hat{\rho} \kappa^{2}}{(s+i \Omega)^{2}-\kappa^{2}}+\left(1-k_{c}^{2}\right) \sigma_{w_{2}}^{2} & k_{c} \frac{\sigma_{u}^{2}+1}{\hat{\delta}} \frac{s+\hat{\delta} \kappa+i \Omega}{s+\kappa+i \Omega} \\
		k_{c} \frac{\sigma_{u}^{2}+1}{\hat{\delta}} \frac{s-\hat{\delta} \kappa+i \Omega}{s-\kappa+i \Omega} & \sigma_{u}^{2}+2
	\end{array}\right] .
\end{aligned}
\end{equation*}
It was shown in \cite[Proposition 1]{Ugrinovskii2024a} that the system satisfies Assumption \ref{ass1} when
\begin{equation}\label{eq80}
	\lambda^{2}>\frac{k_{c}^{2}\left(1+\sigma_{u}^{2}\right)^{2}}{k_{c}^{2} \mu^{2} \hat{\rho}+\left(1-k_{c}^{2}\right) \sigma_{w_{2}}^{2}}-\left(\sigma_{u}^{2}+2\right).
\end{equation}
In this case, the following constants are well defined,
\begin{equation*}
	\alpha_{1}=\left(\frac{\mu_{2}}{\hat{\delta}^{2}\left(\sigma_{u}^{2}+2+\lambda^{2}\right)}\right)^{1 / 2}, \quad \beta_{1}=\left(\frac{\mu_{1}}{\mu_{2}}\right)^{1 / 2},
\end{equation*}
where
\begin{equation*}
	\begin{aligned}
		\mu_{1}= & \left(k_{c}^{2} \mu^{2} \hat{\rho}+\left(1-k_{c}^{2}\right) \sigma_{w_{2}}^{2}\right)\left(\sigma_{u}^{2}+2+\lambda^{2}\right) \\
		& -k_{c}^{2}\left(1+\sigma_{u}^{2}\right)^{2}>0, \\
		\mu_{2}= & \left(k_{c}^{2} \mu^{2}+\left(1-k_{c}^{2}\right) \sigma_{w_{2}}^{2}\right) \hat{\delta}^{2}\left(\sigma_{u}^{2}+2+\lambda^{2}\right) \\
		& -k_{c}^{2}\left(1+\sigma_{u}^{2}\right)^{2}>0.
	\end{aligned}
\end{equation*}
Also, let
\begin{equation*}
\alpha_{2}=\frac{k_{c}\left(1+\sigma_{u}^{2}\right)}{\hat{\delta}\left(\sigma_{u}^{2}+2+\lambda^{2}\right)^{1 / 2}}.
\end{equation*}
It is readily verified that the transfer function
\begin{equation}
\Upsilon_{\lambda}(s)=\left[\begin{array}{cc}
	\alpha_{1} \frac{s+\beta_{1} \hat{\delta} \kappa+i \Omega}{s+\kappa+i \Omega} & \alpha_{2} \frac{s+\hat{\delta} \kappa+i \Omega}{s+\kappa+i \Omega}  \\
	0 & \sqrt{\sigma_{u}^{2}+2+\lambda^{2}}
\end{array}\right]
\end{equation}
is a spectral factor of $\Phi_{\lambda}$. The parameters of its minimal realization \eqref{eq24} are chosen to be
\begin{align}
	A_{\lambda} & =-(\kappa+i \Omega), \;\;B_{\lambda}=\left[\begin{array}{ll}
		\alpha_{1}(\beta_{1} \hat{\delta}-1) \kappa & \alpha_{2}(\hat{\delta}-1) \kappa
	\end{array}\right], \nonumber\\
	C_{1, \lambda} & =1, \quad C_{2, \lambda}=0, \nonumber\\
	D_{1, \lambda} & =\left[\begin{array}{ll}
		\alpha_{1} & \alpha_{2}
	\end{array}\right], \quad D_{2, \lambda}=\left[\begin{array}{ll}
		0 & \sqrt{\sigma_{u}^{2}+2+\lambda^{2}}
	\end{array}\right]. 
\end{align}
Assumption \ref{ass2}(i) is satisfied in this case, since $\alpha_{1}>0$, $\alpha_{2}>0$. Condition (iii) of Assumption \ref{ass2} is also satisfied since the first matrix in \eqref{eq52} is a scalar and the second matrix is a row matrix of dimensions $1 \times 2$. The second condition of Assumption \ref{ass2} was validated numerically and was found to hold as well. Finally, we note that the pair \eqref{eq61} is detectable with any $\lambda \geq 0, \gamma>0$ since $\operatorname{Re} A_{\lambda}<0$ and $C_{1, \lambda}=1$.

\begin{figure}
	\centering
	\includegraphics[width=1\columnwidth]{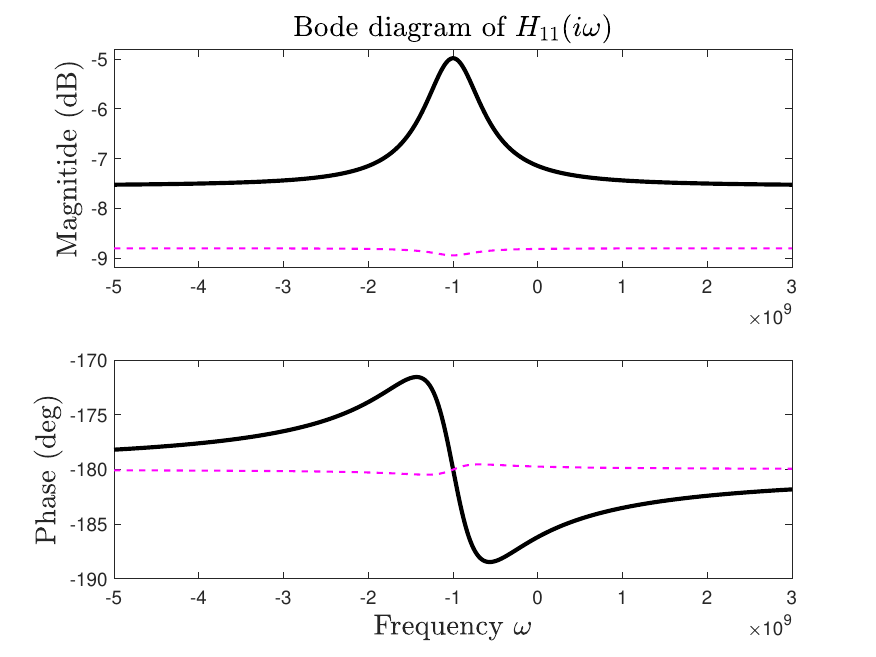}
	{\caption{Bode plots of the transfer function $H_{11}(s)$ in equation \eqref{eq83} (the solid black line) and the corresponding transfer function in \cite{Ugrinovskii2024a} (the dashed magenta line) for $\sigma_{w_2}^2=3$.}\label{fig4}}
\end{figure}

To evaluate the efficacy of the proposed method, we selected the same numerical values for the parameters $\sigma_{u}^{2}, \sigma_{w_{1}}^{2}, k$, $\kappa$, and $\Omega$ as those used in the example in \cite{Ugrinovskii2024a}: $\sigma_{u}^{2}=0.1$, $\sigma_{w_{1}}^{2}=0.2, \sigma_{w_{2}}^{2}=3, k=0.4, k_{c}=1 / \sqrt{2}, \kappa=5 \times 10^{8}$, $\Omega=10^{9}$. Also, we chose the same value of $\gamma^{2}=1.9448$ and found that Assumption \ref{ass2}(ii) was satisfied with $\lambda=0$. Condition \eqref{eq80} is also satisfied with these parameters. Then we applied Algorithm 1 to obtain the corresponding coherent equalizer transfer function $H(s)$:
\begin{equation}\label{eq83}
	\begin{aligned}
		H_{11}(s)=&-0.41494 \frac{s+5 \times 10^{8}+1 \times 10^{9} i}{s+3.718 \times 10^{8}+1 \times 10^{9} i},\\
		H_{12}(s)=&-0.90985 \frac{s+3.392 \times 10^{8}+1 \times 10^{9} i}{s+3.718 \times 10^{8}+1 \times 10^{9} i},\\
		H_{21}(s)=&0.90985 \frac{s-3.392 \times 10^{8}+1 \times 10^{9} i}{s+3.718 \times 10^{8}+1 \times 10^{9} i},\\
		H_{22}(s)=&-0.41494 \frac{s-5 \times 10^{8}+1 \times 10^{9} i}{s+3.718 \times 10^{8}+1 \times 10^{9} i}.
	\end{aligned}
\end{equation}
In Step 5 of the algorithm, the unitary transfer function $U(s)=\frac{s-3.392 \times 10^{8}+1 \times10^{9} i}{s+3.392 \times 10^{8}+1 \times10^{9} i}$ was used. The Bode plots of $H_{11}$ and the corresponding transfer function obtained in \cite{Ugrinovskii2024a} for this system are shown in Fig.~\ref{fig4}. 
Ref.~\cite{Ugrinovskii2024a} considered a low signal-to-noise ratio scenario, where the resulting transfer function is nearly constant, as illustrated in Fig.~\ref{fig4}. This behavior suggests that, in highly noisy environments, minimizing the PSD in \eqref{eq1} is effectively achieved by uniformly suppressing the magnitude across all frequencies of the input signal, resulting in a constant-like transfer function when using the algorithm from~\cite{Ugrinovskii2024a}.

\begin{figure}
	\centering
	\includegraphics[width=1\columnwidth]{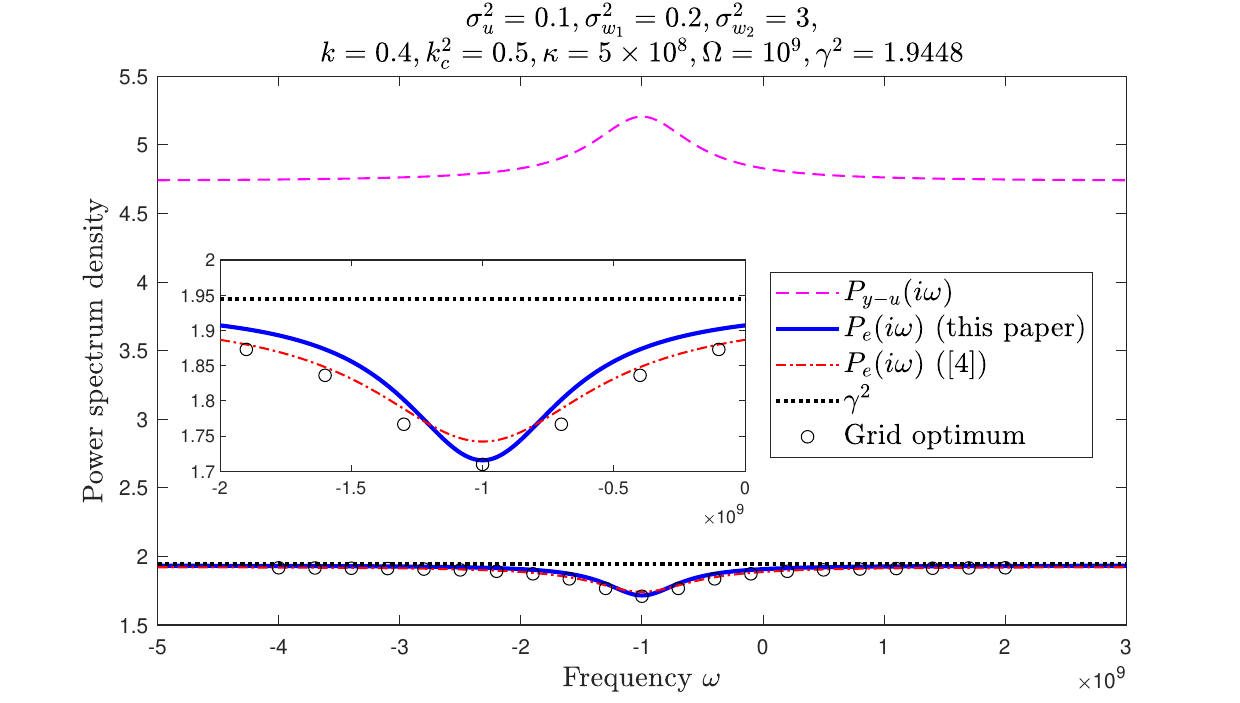}
	{\caption{Power spectrum density $P_e(i\omega)$ for the system with the equalizer \eqref{eq83} (the solid blue line). Also shown in the figure are the power spectrum density $P_{y-u}(i\omega)$ of the difference $y-u$ (the dashed magenta line), the power spectrum density $P_e(i\omega)$ for the system with the equalizer obtained in \cite{Ugrinovskii2024a}  (the dash-dotted red line) for this $\gamma^2$, and the values obtained from the semidefinite program \eqref{eq74} (the circles).}\label{fig5}}
\end{figure}

The plot of the error power spectrum density $P_{e}(i \omega)$ for the resulting equalizer shown in Fig. \ref{fig5} confirms that with this equalizer, $P_{e}(i \omega)<\gamma^{2}I_{n}$. Also, Fig. \ref{fig5} compares the power spectrum density $P_{e}$ with the power spectrum density of the difference between the channel input and output $y-u$, i.e., when the channel output is not equalized. We see that the output of the equalizer represents $u$ with a higher mean-square fidelity than the output $y$ of the channel.

Also Fig. \ref{fig5} shows the power spectrum density of the equalization error achieved using the equalizer obtained in \cite{Ugrinovskii2024a}. As expected, the two equalizers perform quite similarly, given that the chosen $\gamma^{2}$ is very close to the optimal value \eqref{eq2} in this example, according to \cite{Ugrinovskii2024a}. To confirm that and to demonstrate application of Theorem \ref{th3}, Step 6 of Algorithm 1 was carried out using a grid of
$21$ frequency points $\omega_{l}$ evenly distributed in the interval $\left[-4 \times 10^{9}, 2 \times 10^{9}\right]$. The corresponding value $\nu_{\bar{\omega}}^{2}$ was found to be $\nu_{\bar{\omega}}^{2} \approx 1.9191$. This confirms that the chosen value of $\gamma^{2}=1.9448$ is close to $\inf \sup _{\omega} \bar{\boldsymbol{\sigma}}\left(P_{e}(i \omega)\right)$ within a $1.34 \%$ margin which explains the similarity in performance between the equalizer \eqref{eq83} and that obtained in \cite{Ugrinovskii2024a}.

\begin{figure}
	\centering
	\includegraphics[width=1\columnwidth]{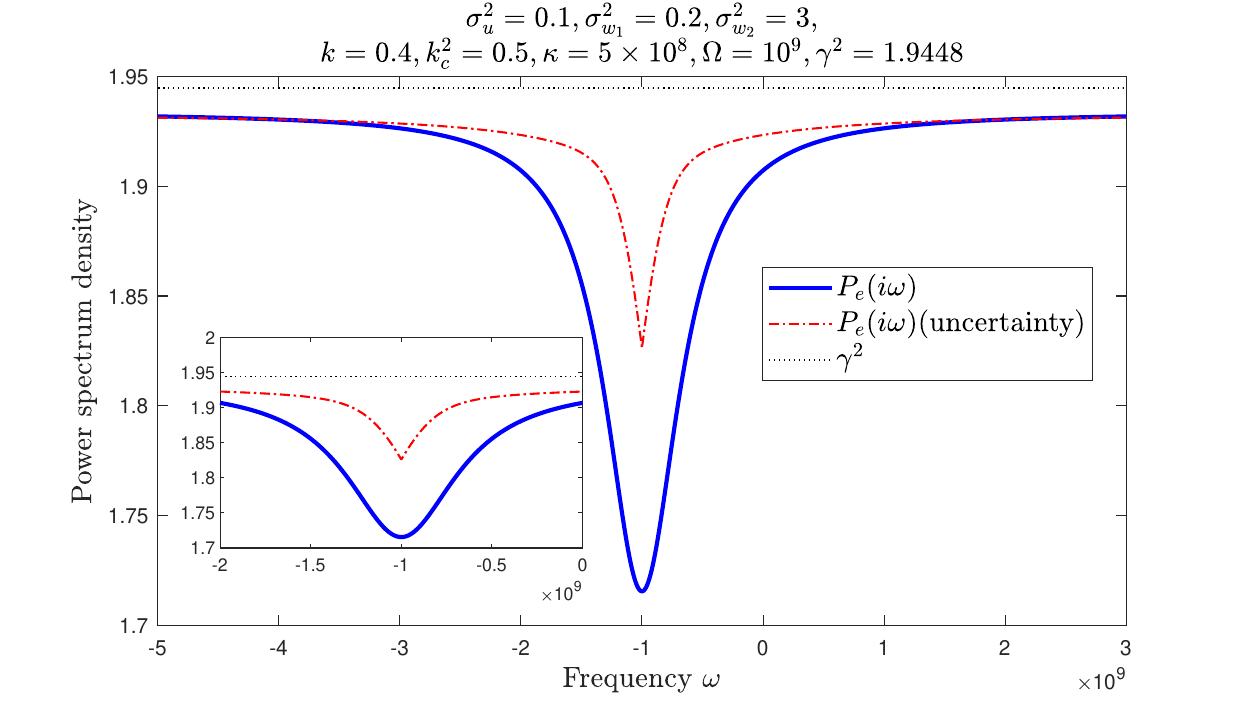}
	\caption{Power spectrum density \( P_e(i\omega) \) for the system with the equalizer given in~\eqref{eq83} (solid blue line). Also shown is \( P_e(i\omega) \) under Hamiltonian uncertainty, where \( \Omega \) varies within the range \( [6\times10^8, 14\times10^8] \) (dash-dotted red line), with \( \gamma^2 = 1.9448 \).\label{un}}
\end{figure}

To evaluate robustness, we introduce uncertainty in the Hamiltonian of the optical cavity, varying \( \Omega \) within the range \( [6\times10^8, 14\times10^8] \) with a grid step of \( 1.6\times10^7 \). Using the equalizer in \eqref{eq83}, we compute the maximum PSD across this range, with \( \gamma^2 = 1.9448 \). The results, presented in Fig.~\ref{un}, show that although Hamiltonian uncertainty increases the PSD compared to the nominal case, it remains below the bound \( \gamma^2 = 1.9448 \), thereby demonstrating the robustness of the equalizer.

Next, we repeated simulations with $\sigma_{w_{2}}^{2}=0.2$. For this value of $\sigma_{w_{2}}^{2}$, the method of \cite{Ugrinovskii2024a} fails to produce a physically realizable equalizer. However, using the method proposed in this paper we were able to obtain the smallest (within an absolute tolerance of $10^{-3}$ ) $\gamma^{2}$ for which the conditions of Theorem \ref{th2} were satisfied when $\sigma_{w_{2}}^{2}=0.2$. For this, $\lambda^{2}=1.9343$ was chosen using \eqref{eq80}. Then Steps 1-3 of Algorithm 1 were repeated and $\gamma^{2}$ was adjusted using the bisection method until the desired precision was reached. The resulting optimized $\gamma^{2}$ was found to be approximately equal to $1.2506$ while the semidefinite program \eqref{eq74} produced the lower bound $\nu_{\bar{\omega}}^{2} \approx 1.2068$,
using the same frequency grid. According to Theorem \ref{th3}, this indicates that the accuracy of approximating \eqref{eq2} with $\gamma^{2}=1.2506$ is within $4 \%$. The equalizer transfer function $H(s)$ was found to have the elements
\begin{equation}\label{eq84}
	\begin{aligned}
		H_{11}(s)=&-0.95793 \frac{s+5 \times 10^{8}+1\times10^{9} i}{s+8.26 \times 10^{8}+1\times 10^{9} i},\\
		H_{12}(s)=&-0.28701 \frac{s+2.345 \times 10^{9}+1\times10^{9} i}{s+8.26 \times 10^{8}+1\times 10^{9} i},\\
		H_{21}(s)=&0.28701 \frac{s-2.345 \times 10^{9}+1\times10^{9} i}{s+8.26 \times 10^{8}+1\times 10^{9} i},\\
		H_{22}(s)=&-0.95793 \frac{s-5 \times 10^{8}+1\times10^{9} i}{s+8.26 \times 10^{8}+1\times 10^{9} i}.
	\end{aligned}
\end{equation}
Note that $\left\|H_{11}\right\|_{\infty} \approx 0.9579$, i.e., $H_{11}(s)$ is contractive as expected, however it has a larger gain compared with the equalizer found for $\sigma_{w_{2}}^{2}=3$.

\begin{figure}
	\centering
	\includegraphics[width=1\columnwidth]{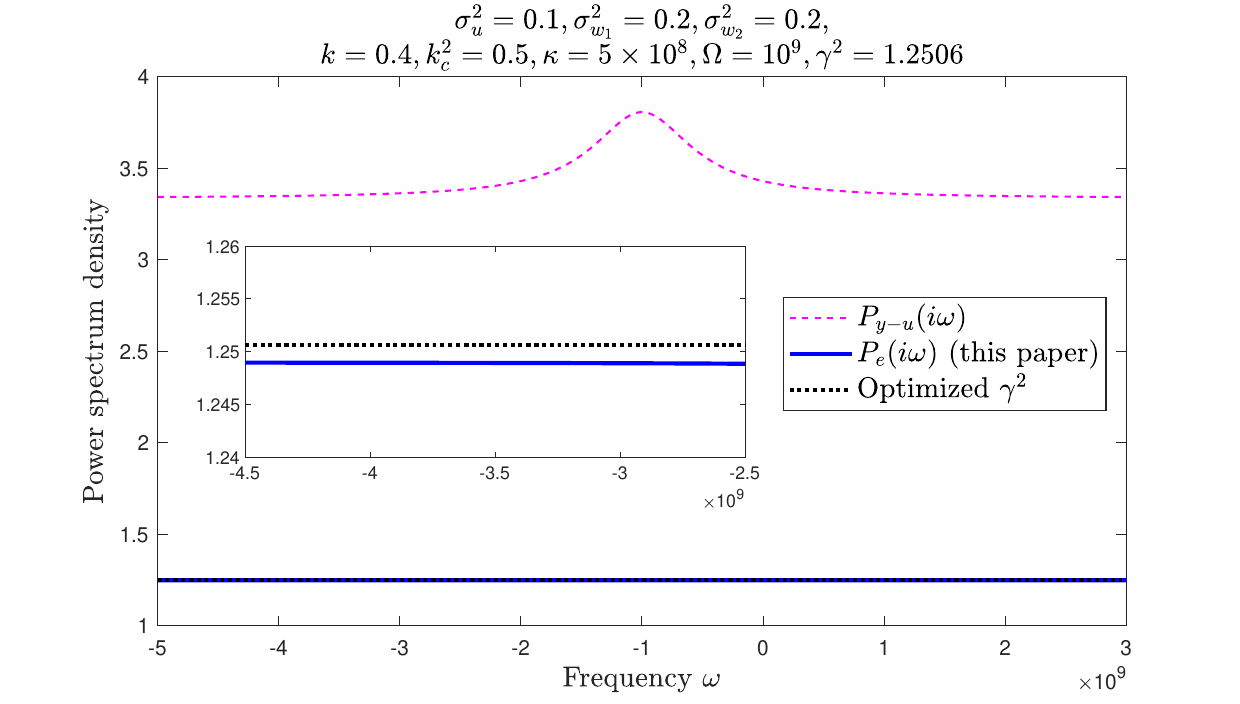}
	{\caption{Power spectrum density $P_e$ of the equalizer (the solid blue line) corresponding to $\gamma^2 = 1.2506$. Also shown in the figure is the power spectrum density $P_{y-u}$ of the difference $y-u$ (the dashed magenta line).}\label{fig6}}
\end{figure}

The plot of the error power spectrum density $P_{e}(i \omega)$ for the resulting equalizer is shown in Fig. \ref{fig6}. As expected, the found $\gamma^{2}$ bounds $P_{e}(i \omega)$ from above. As in the previous case, the power spectrum density $P_{y-u}$ of the difference between the channel input $u$ and output $y$ is substantially greater.

\begin{figure}
	\centering
	\includegraphics[width=1\columnwidth]{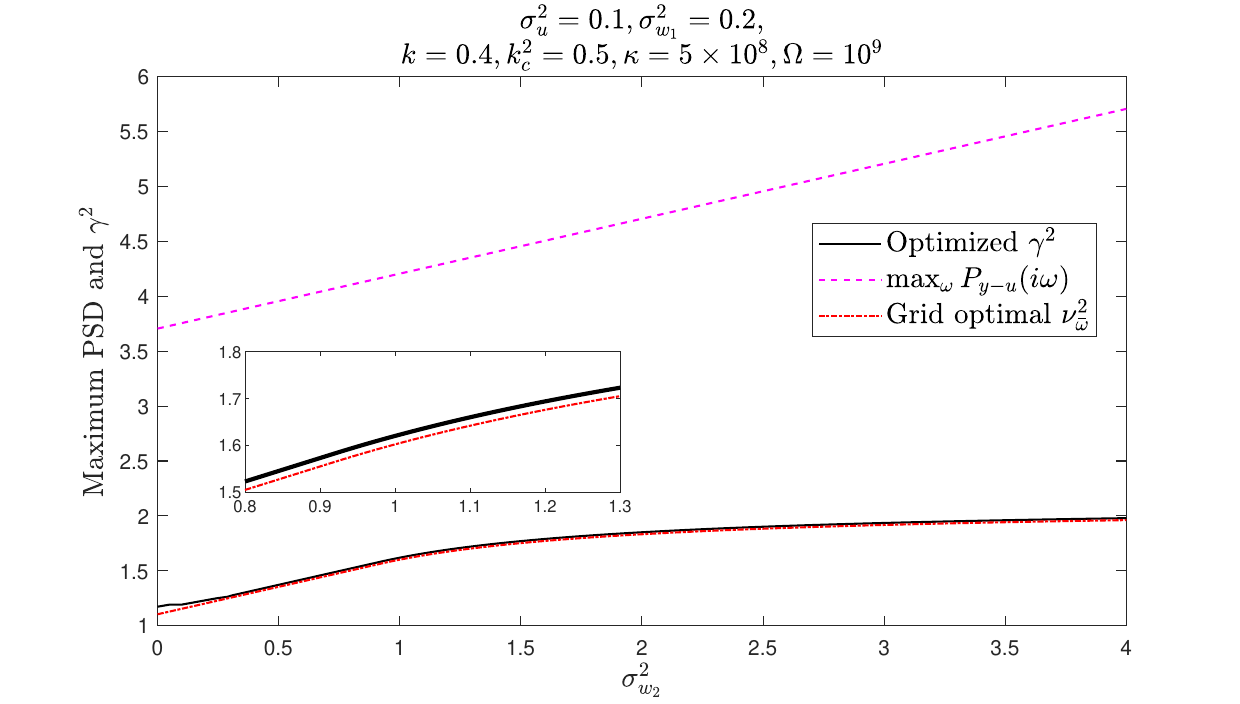}
	{\caption{Optimized $\gamma^2$ (the solid black line), the maximal value of $P_{y-u}(i\omega)$
		 (the dashed magenta line) and $\nu_{\bar\omega}^2$ for a range of $\sigma_{w_2}^2$.}\label{fig7}}
\end{figure}

We also repeated simulations for a range of $\sigma_{w_{2}}^{2} \in[0,4]$. For each $\sigma_{w_{2}}^{2}$, the smallest $\gamma^{2}$ was computed (within a precision of $10^{-3}$) for which the conditions of Theorem \ref{th2} were satisfied. Also, we used the same grid of 21 frequency points to compute the corresponding values $\nu_{\bar{\omega}}^{2}$ of the semidefinite program \eqref{eq74}. Fig. \ref{fig7} shows the graph of the obtained optimized $\gamma^{2}$. We observe that the mean-square performance of the equalizers obtained via Algorithm 1 follows closely the values of the semidefinite program \eqref{eq74}. Thus in this example, for each $\sigma_{w_{2}}^{2}$ within the considered interval, the method in this paper produces a physically realizable equalizer $H(s)$ whose equalization performance approximates the optimal performance objective \eqref{eq2} with reasonable accuracy. Although the method in \cite{Ugrinovskii2024a} led to equalizers with a theoretically exact optimal performance, such equalizers could only be derived when $\sigma_{w_{2}}^{2}$ was sufficiently large.

\begin{figure}
	\centering
	\includegraphics[width=0.8\columnwidth]{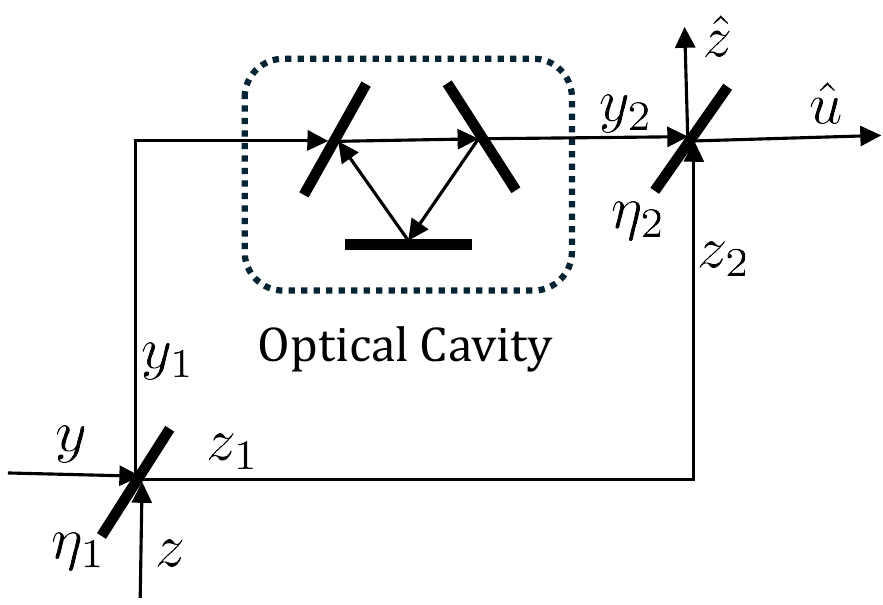}
	{\caption{A cavity and beam splitters realization of the equalizer \eqref{eq83}.}\label{fig8}}
\end{figure}

We finish this example by noting that the equalizers $H(s)$ in equations \eqref{eq83} and \eqref{eq84} can be realized as a quantum optical system consisting of an optical cavity and two beam splitters; see Fig. \ref{fig8}. The transfer function of the optical cavity in this realization is
\begin{equation*}
	H_{c}(s)=\frac{s-\kappa_{1}+i \Omega}{s+\kappa_{1}+i \Omega},
\end{equation*}
where $\kappa_{1}=-\operatorname{Re} p_{1}, p_{1}$ is the pole of $H(s)$. The beam splitters' operators are
\begin{equation*}
\left[\begin{array}{l}
	y_{1} \\
	z_{1}
\end{array}\right]=\left[\begin{array}{cc}
	\xi_{1} & \eta_{1} \\
	\eta_{1} & -\xi_{1}
\end{array}\right]\left[\begin{array}{l}
	y \\
	z
\end{array}\right], \quad\left[\begin{array}{l}
	\hat{u} \\
	\hat{z}
\end{array}\right]=\left[\begin{array}{cc}
	\eta_{2} & \xi_{2} \\
	\xi_{2} & -\eta_{2}
\end{array}\right]\left[\begin{array}{l}
	y_{2} \\
	z_{2}
\end{array}\right].
\end{equation*}
To obtain the values of the coefficients $\eta_{1}, \xi_{1} \triangleq \sqrt{1-\eta_{1}^{2}}$, $\eta_{2}, \xi_{2} \triangleq \sqrt{1-\eta_{2}^{2}}$, it is convenient to write $H_{11}(s)$ as
\begin{equation}
	H_{11}(s)=a \frac{s+b \kappa_{1}+i \Omega}{s+\kappa_{1}+i \Omega}.
\end{equation}

Since the auxiliary synthesis problem in Definition \ref{def2} restricts $H_{11}(s)$ to lie in the class of contractive transfer functions, then $|a|<1$. Also, in \eqref{eq83} $\kappa_{1}=3.7184 \times 10^{8}$ and $b=1.3447>1$, while in \eqref{eq84} $\kappa_{1}=8.2603 \times 10^{8}$ and $b=0.6053<1$.

In the case where $b<1$, the implementation of $H(s)$ as a system in Fig. \ref{fig8} was discussed in \cite{Ugrinovskii2024a}. The values of the coefficients $\eta_{1}$, $\xi_{1} \triangleq \sqrt{1-\eta_{1}^{2}}$, $\eta_{2}$, $\xi_{2} \triangleq \sqrt{1-\eta_{2}^{2}}$ are expressed in terms of $a, b$, as
\begin{align}
	& \eta_{1}=-\sqrt{\frac{1+a^{2} b-\sqrt{\left(1-a^{2} b^{2}\right)\left(1-a^{2}\right)}}{2}}, \nonumber\\
	& \xi_{1}=\sqrt{1-\eta_{1}^{2}}=\sqrt{\frac{1-a^{2} b+\sqrt{\left(1-a^{2} b^{2}\right)\left(1-a^{2}\right)}}{2}}, \nonumber\\
	& \eta_{2}=-\sqrt{\frac{1-a^{2} b-\sqrt{\left(1-a^{2} b^{2}\right)\left(1-a^{2}\right)}}{2}}, \nonumber\\
	& \xi_{2}=\sqrt{1-\eta_{2}^{2}}=\sqrt{\frac{1+a^{2} b+\sqrt{\left(1-a^{2} b^{2}\right)\left(1-a^{2}\right)}}{2}}. \label{eq86}
\end{align}
The same implementation can be used when $b>1$, and $a^{2} b^{2}<1$, which is the case in \eqref{eq83}. However, in this case the second beam splitter must be tuned differently. Namely, the values of the coefficients $\eta_{1}, \xi_{1}$ are the same as in \eqref{eq86}, but $\eta_{2}$ and $\xi_{2}$ must be set to
\begin{equation*}
\begin{aligned}
	& \eta_{2}=\sqrt{\frac{1-a^{2} b-\sqrt{\left(1-a^{2} b^{2}\right)\left(1-a^{2}\right)}}{2}} \\
	& \xi_{2}=\sqrt{1-\eta_{2}^{2}}=\sqrt{\frac{1+a^{2} b+\sqrt{\left(1-a^{2} b^{2}\right)\left(1-a^{2}\right)}}{2}}
\end{aligned}
\end{equation*}
Note that since $a^{2} b^{2}<1$ in this case, the above parameters remain real.

\section{Conclusions}\label{sec6}
The paper has developed a new methodology for the synthesis of completely passive mean-square near-optimal coherent equalizers for quantum communication channels. We have shown that the synthesis problem reduces to an auxiliary two-disk $H_{\infty}$ control problem for an associated classical control system with disturbance feedforward. A solution to this auxiliary problem has been developed which draws on the richness of the set of $H_{\infty}$ controllers.

The paper has integrated the aforementioned solution to the auxiliary problem into an algorithm for the design of physically realizable equalizers. We have been able to circumvent more restrictive conditions required for this in the recent work \cite{Ugrinovskii2024a}. As a result, the domain of applicability of the proposed method has been expanded substantially, as the benchmark example in Section \ref{sec5} has demonstrated. We have observed that when both methods are applicable, the method developed in this paper yields a physically realizable transfer function whose equalization performance matches closely the performance of optimal equalizers from \cite{Ugrinovskii2024a}. In addition, the proposed method has been shown to work in low noise scenarios where the previous method could not be applied or failed to produce a solution.

Our current work is limited to passive channels and passive equalizers. In future research, we plan to extend this framework to incorporate active channels and equalizers, thereby providing a more general and versatile solution. Preliminary studies in this direction have been reported in~\cite{valcdc2}.

This work also assumes an accurate model of the quantum channel. As part of our future efforts, we aim to develop a general framework for robust quantum coherent equalization to systematically address parameter perturbations and enhance performance under model uncertainties.
Furthermore, there remains a gap between the optimal value of \( \gamma \) obtained using our method and the lower bound derived from semidefinite programming. Future work will focus on developing an improved framework, beyond the current two-disk formulation, to achieve a tighter approximation and further reduce this gap.
Finally, for multi-mode quantum systems, solving LMIs can be computationally intensive. We intend to explore more efficient algorithms and numerical methods \cite{lmia1,lmia2}—such as advanced convex relaxation techniques, model reduction strategies, and customized iterative solvers—to improve the computational scalability of our approach for large-scale systems.

\begin{ack}
	The authors gratefully acknowledge inspiring discussions with Prof. Matthew R. James on the topic of coherent equalization. The authors are also grateful to Prof. James for his valuable comments on the manuscript.
\end{ack}

\bibliographystyle{ieeetr}        % Include this if you use bibtex 
\bibliography{basic}           % and a bib file to produce the 
                                 % bibliography (preferred). The
                                 % correct style is generated by
                                 % Elsevier at the time of printing.

\end{document}